\documentclass[twocolumn]{aa}
\usepackage{graphicx}
\usepackage{txfonts}
\usepackage{lscape}
\usepackage{rotating}
\usepackage{longtable}

\newcommand\4{{\footnotesize IV}}
\newcommand\3{{\footnotesize III}}
\newcommand\2{{\footnotesize II}}

\newcommand\lam{{$\lambda$}}
\newcommand\kms{$\rm{km s^{-1}}$}
\begin{document}
   \title{Physical Parameters 
and Wind Properties of Galactic Early B Supergiants}

\titlerunning{Early B supergiants}

   \author{Paul A. Crowther\inst{1} 
\and          Daniel J. Lennon\inst{2}
          \and Nolan R. Walborn\inst{3}
}

   \offprints{P. A. Crowther (Paul.Crowther@sheffield.ac.uk)}

   \institute{ Department of Physics and
Astronomy, University of Sheffield, Hicks Building, Hounsfield Rd,
Shefffield, S3 7RH,  UK
          \and
Isaac Newton Group, Apartado 321, 38700 Santa Cruz de La Palma, Canary Islands, Spain
         \and
Space Telescope Science Institute, 3700 San Martin Drive, Baltimore MD 21218, USA             
}
\date{Received March 2005/Accepted}

   \abstract{We present optical 
studies of the physical and wind properties, plus 
CNO chemical abundances, of 25 O9.5--B3 Galactic supergiants. 
We employ non-LTE, line blanketed, extended 
model atmospheres, which provide a modest downward revision in the effective 
temperature scale of early B supergiants of up to 1--2kK relative to previous 
non-blanketed results. The so-called `bistability jump' at B1 ($T_{\rm eff} \sim$ 21kK)
from Lamers et al. is rather a more gradual trend (with large scatter) 
from  $v_{\infty}/v_{\rm esc}\sim$3.4 for
B0--0.5 supergiants above 24kK to  $v_{\infty}/v_{\rm esc}\sim$2.5 for B0.7--1  
supergiants with 20kK$\leq T_{\rm eff} \leq$24kK, and  $v_{\infty}/v_{\rm esc}\sim$1.9
for B1.5--3 supergiants below 20kK. This, in part, explains the break in observed UV spectral 
characteristics between B0.5 and B0.7  subtypes as discussed by Walborn et al. We compare derived 
(homogeneous) wind densities with recent results for Magellanic Cloud 
B supergiants and generally confirm theoretical expectations for stronger winds amongst Galactic 
supergiants. However, winds are substantially weaker than predictions from current radiatively 
driven wind theory, especially at mid-B subtypes, 
a problem which is exacerbated if winds are already clumped in the  H$\alpha$ 
line forming region. In general,
CNO elemental abundances reveal strongly  processed
material at the surface of Galactic B supergiants, with mean N/C and N/O
abundances 10 and 5 times higher than the Solar value, 
respectively, with HD~2905 (BC0.7\,Ia) indicating the lowest degree of 
processing in our sample, and HD~152236 (B1.5\,Ia$^{+}$) the highest.
   }

   \maketitle
%

\section{Introduction}\label{sect1}

Early-type stars possess powerful winds that provide significant
feedback to the local ISM (e.g. Smith 2005). Significant progress has recently been
made towards an improved understanding of the physical parameters and
winds of massive O stars  and their immediate descendants, OBA supergiants. 
In particular, the availability of modern spectral synthesis codes allowing for
line blanketing and stellar winds  together with high-resolution
instruments on 4--8m telescopes permits detailed studies of OB stars
in the Magellanic Clouds (e.g. Crowther et al. 2002; Hillier et al. 2003;
Trundle et al. 2004). 
A primary motive behind
such studies is the investigation of how winds in the low-metallicity
LMC and SMC differ from those in the Milky Way. 

For O stars, 
Herrero et al. (2003) and Repolust et al. (2004) provide suitable 
Galactic comparison stars, whilst one currently has to rely on 
results from Kudritzki et al. (1999) for B supergiants, for which
the temperature scale of McErlean et al. (1999) was
adopted.  This is one of the principal reasons behind the present study,
namely a determination of the wind properties and temperature scale
of B0--3 supergiants using contemporary 
line blanketed techniques and optical spectroscopy. Although the range
in spectral type is narrow, the effective temperatures of B0--3 supergiants
span almost a factor of {\it two}, from 17 to 32kK (McErlean et al. 1999).
The present optical study of ionization in early B supergiants 
complements the recent UV wind ionization investigation by Prinja et al. (2005).
Further, there is now considerable qualitative evidence supporting CNO-processed
material in the atmospheres of massive stars (Herrero \& Lennon 2004), although quantitative
results for B supergiants -- for comparison with recent evolutionary models
allowing for rotation (Meynet \& Maeder 2000) -- are generally lacking.

\begin{table*}
\caption{Program Stars - observations are from either the 1m JKT (Lennon
et al. 1992), 2.5m INT or 4.2m WHT (northern) or 1.5m CTIO 
(southern).  Two
estimates of interstellar reddenings are given, based on intrinsic colours adopted
from Schmidt-Kaler (1982) and from theoretical continuum fits to UV spectrophotometry 
and optical photometry (no spectrophotometry was available for HD~194279 or HD~148688). 
The majority of our targets are probable cluster/association members,
whilst absolute magnitudes are adopted for the remainder, based on average values of
Magellanic Cloud members from Fitzpatrick (1991),  Parker (1993) and
 Massey et al. (1995), except  that  HD~190603 is set at $M_{\rm V}=-7.5$ mag. 
References to  cluster/association distances are: 
(a) Brown et al. (1994); (b) Blaha \& Humphreys (1989); 
(c) Baume et al. 1999;
(d) Baume et al (2003); (e) Garmany \& Stencel (1992); (f) Humphreys (1978).\label{table1}}
\begin{tabular}{
r@{\hspace{2mm}}c@{\hspace{2mm}}l@{\hspace{2mm}}
l@{\hspace{2mm}}r@{\hspace{2mm}}
c@{\hspace{2mm}}
l@{\hspace{2mm}}
l@{\hspace{2mm}}
r@{\hspace{2mm}}
c@{\hspace{2mm}}
c@{\hspace{2mm}}
c@{\hspace{2mm}}
c@{\hspace{2.5mm}}l
}
\noalign{\hrule\smallskip}
HD & Name & Sp Type & V & B-V & \multicolumn{2}{c}{E(B-V)}  &Assoc.& 
DM&Ref &  $M_{\rm V}$ & $\Delta M_{\rm V}$ & Obs\\
   &      &         &mag& mag & (B-V)$_{0}$     & $UV$      & mag     &    & mag & mag \\
\noalign{\hrule\smallskip}
30614 & $\alpha$ Cam & O9.5\,Ia &4.29&+0.02 & 0.29& 0.32 &          &10.0 & & --6.6 & $\pm$ 0.5 & WHT\\  
37128 & $\epsilon$ Ori & B0\,Ia &1.70&--0.19& 0.05&0.06 & Ori OB1b & 7.8 &a& --6.3 & $\pm$ 0.5 & WHT/INT\\ 
91969 &                & B0\,Ia &6.52&+0.00 & 0.24&0.26& NGC3293 &12.2 &d&--6.4 & $\pm$0.3 & CTIO \\ 
94909 &                & B0\,Ia &7.34&+0.48 &0.72 &0.75$\ddag$&         &11.5 & &--6.4 & $\pm$0.5 & CTIO\\ 
122879&                & B0\,Ia &6.42&+0.12 &0.36 &0.37&         &11.7 & &--6.4 & $\pm$0.5 & CTIO \\ 
38771 & $\kappa$ Ori & B0.5\,Ia&2.05&--0.18 &0.04 &0.07 & Ori OB1c & 8.0 &a&--6.1 &$\pm$0.5 & WHT/INT\\ 
115842&              &B0.5\,Ia &6.02&+0.29 &0.51  &0.57$\ddag$&         & 11.3& &--6.9 & $\pm$0.5 & CTIO\\ 
152234&        &B0.5\,Ia (N wk)&5.44&+0.20 &0.42  &0.45& Sco OB1 & 11.5&c&--7.4 & $\pm$0.5 & CTIO\\ 
2905&$\kappa$ Cas&BC0.7\,Ia &4.16&+0.14 & 0.35  & 0.39& Cas OB14& 10.2 &b&--7.1 & $\pm$0.3 & WHT/INT\\ 
91943 &            & B0.7\,Ia &6.73&+0.07 & 0.28 &0.22$\ddag$& NGC~3293& 12.2 &d&--6.3  & $\pm$0.3 & CTIO \\ 
152235&      & B0.7\,Ia (N wk)&6.34&+0.51 & 0.72 &0.75$\ddag$& Sco OB1 & 11.5 &c&--7.4 & $\pm$0.5 & CTIO \\ 
154090&            & B0.7\,Ia &4.87&+0.26 & 0.47 &0.49$\ddag$&         & 10.2 & &--6.8 & $\pm$0.5 & CTIO \\ 
13854&             & B1\,Iab& 6.48&+0.28& 0.47  &0.55& Per OB1 & 11.8 &e&--6.8 & $\pm$0.5 & WHT/INT\\ 
91316& $\rho$  Leo & B1\,Iab (N str)&3.85 &--0.14& 0.05 & 0.06  &         & 10.5 & &--6.8 & $\pm$0.5 & CTIO \\ 
148688&            & B1\,Ia  &5.33 &+0.33& 0.52  &0.52$\ddag$&        & 10.5 & &--6.8 & $\pm$0.5 & CTIO \\
14956 &            & B1.5\,Ia &7.19 &+0.72 &0.90  &0.95$\ddag$& Per OB1 & 11.8 &e& --7.4  & $\pm$0.5 & WHT/INT\\ 
152236& $\zeta^{1}$ Sco & B1.5\,Ia$^{+}$ &4.73 &+0.48 & 0.66 &0.69&Sco OB1& 11.5&c& --8.8 & $\pm$0.5 & CTIO\\
190603& & B1.5\,Ia$^{+}$    &5.65 &+0.54 &0.72  & 0.80$\ddag$&  & 10.9 & &--7.5 & $\pm$1.0 & WHT/INT\\
14143 &        & B2\,Ia  & 6.66 &+0.50 &0.67 &0.70& $h$ Per  & 11.8  &e&--7.2  & $\pm$0.3 & WHT/INT\\ 
14818 & 10 Per & B2\,Ia & 6.25 &+0.30 &0.47 &0.52& Per OB1 & 11.8 &e&--7.0 & $\pm$0.5 & WHT/INT\\ 
41117 & $\chi^2$ Ori & B2\,Ia & 4.63 & +0.28 & 0.45 & 0.48&Gem OB1 & 10.9 & b&--7.6 & $\pm$0.3 & WHT\\ 
194279&        & B2\,Ia  & 7.05 &+1.02 &1.19  &1.26$\ddag$&Cyg OB9  & 10.4  &f&--7.0   & $\pm$0.3 & WHT/INT\\ 
198478 & 55  Cyg& B2.5\,Ia & 4.86& +0.42 & 0.57 &0.59&Cyg OB7 & 9.5 &f&--6.4 & $\pm$0.3& WHT\\ 
14134 &          & B3\,Ia & 6.55&+0.45  & 0.58 &0.65$\ddag$&$h$ Per& 11.8  &e&--7.1 & $\pm$0.3 & JKT/INT \\
53138 &$o^2$ CMa & B3\,Ia & 3.01&--0.08 & 0.05 & 0.07&     & 10.2  & &--7.3 & $\pm$0.5 &JKT/INT\\
\noalign{\hrule\smallskip}
\end{tabular}
\newline $\ddag$: UV spectrophotometry absent or poor quality, leading to 
uncertain reddening from this technique.
\end{table*}

A second motive for the present study is an observational
investigation of the so-called `bistability jump'
in the terminal velocities of 
early B supergiants (Lamers et al. 1995).
Walborn \& Nichols-Bohlin (1987)
first remarked upon the UV morphological differences between B0.5\,Ia
and B0.7\,Ia supergiants, in the sense that C\,{\sc ii} $\lambda$1335
and Al\,{\sc iii} $\lambda$1855-63 P Cygni profiles first appear at B0.7,
whilst the Si\,{\sc iv} $\lambda$1393-1402 doublet becomes narrower and 
stronger. In solely two cases, HD\,115842 (B0.5\,Ia) and HD\,91943 (B0.7\,Ia)
are these distinctions reversed (Walborn et al. 1995). 
Pauldrach \& Puls (1990) first described a bi-stability jump with reference to 
the extreme B supergiant P Cygni (B1.5\,Ia$^+$), in the sense that
a small change in physical properties for P Cyg can result in 
dramatic differences in the outer wind, 
either ionized or recombined (Najarro et al. 1997; Crowther 1997). Amongst normal
B supergiants, with much weaker winds, Lamers et al. (1995) demonstrated that
the ratio  $v_{\infty}/v_{\rm esc}$ 
drops steeply from about 2.6 for temperatures above
21kK (B1 subtype), to a value of 1.3 below\footnote{Prinja 
\& Massa (1998) suggested a larger observational spread in 
$v_{\infty}/v_{\rm esc}$ amongst early B supergiants}. 

The empirical B supergiant bistability jump of Lamers et al. has been widely
used in synthetic spectral model calculations,
of application to distant star forming galaxies (e.g. Rix et al. 2004).
Vink et al. (1999, 2000) predicted a jump in mass-loss around 25kK, 
due to the line acceleration of Fe\,{\sc iii}, such that the mass-loss 
increases by a factor of 5 between 27.5kK and 22.5kK,  although the change 
in velocity was not thoroughly explained. Here, we present studies of B 
supergiants to investigate the Lamers et al. bistability jump, and test 
whether the Vink et al. predictions are consistent with observations.

\begin{table*}
\caption[]{Atomic models for early and mid B supergiants. For each ion 
F denotes full levels, S super levels, and T the 
number of bound-bound transitions considered. 
Most ions are common
to all stars, except that high ionization stages are
only  included in early B supergiant models (shown in italics).}
\label{table2}
\begin{center}
\begin{tabular}{l@{\hspace{-5mm}}
c@{\hspace{1mm}}c@{\hspace{1mm}}c
c@{\hspace{1mm}}c@{\hspace{1mm}}c
c@{\hspace{1mm}}c@{\hspace{1mm}}c
c@{\hspace{1mm}}c@{\hspace{1mm}}c
c@{\hspace{1mm}}c@{\hspace{1mm}}c
c@{\hspace{1mm}}c@{\hspace{1mm}}c
c@{\hspace{1mm}}c@{\hspace{1mm}}c}
\noalign{\hrule\smallskip}
Element & 
\multicolumn{3}{c}{I} & 
\multicolumn{3}{c}{II} & 
\multicolumn{3}{c}{III} & 
\multicolumn{3}{c}{IV} & 
\multicolumn{3}{c}{V}  \\
        &F & S & T 
	&F & S & T 
	&F & S & T 
	&F & S & T
	&F & S & T
	\\
\noalign{\hrule\smallskip}

H       & 30 & 20 & 435 \\
He      & 59 & 41 & 590 & 30 & 20 & 435 \\
C       &    &    &     & 53& 30 &  323& 54& 29 & 268 
                        & {\it 18}&{\it 13} & {\it 76}       \\
N       & 22 & 10& 59& 41& 21 &  144& {\it 70} & {\it 34}& 
{\it 430}& \\
O       & 75 & 18 &  450 &  63&  22 & 444& {\it 45} & {\it 25}& {\it 182} \\ 
Mg      &    &    &     & 45 & 18 & 362 \\
Al      &    &    &     & 44 & 26 &  171 & 65 & 21 & 1452 \\
Si      &    &    &     & 62 & 23 &  365  & 45 & 25& 172 &  12 & 8 & 26 \\
S       &    &    &     & 87 & 27 &  786  & 41  & 21& 177& {\it 92}&{\it 37}& 
{\it 708} \\
Ca      &    &    &     & 12 & 7 & 28\\
Fe      &    &    &     & 510 & 100 & 7501 & 607 & 65&  5482& 272&    48 &3113 
& {\it 182}  & {\it 46} &{\it 1781}  \\
\noalign{\hrule\smallskip}
\end{tabular}
\end{center}
\end{table*}

\section{Observations}\label{sect2}

We have selected 25 Galactic early-type 
supergiants in the spectral range O9.5--B3, 75\% of which have established
cluster/association membership, listed in Table~\ref{table1}. In all but two cases, robust
UV wind velocities are available from the literature -- otherwise
average subtype values from Howarth et al. are adopted (HD~19456, 
HD~192479). Cluster or association distance estimates are available for the
majority of the sample, whilst average subtype absolute magnitudes are
adopted for the remainder based on Magellanic Cloud stars (Fitzpatrick 1991,
Parker 1993, Massey et al. 1995). Many of the associations are rather extended
(e.g. Per OB1: Kendall et al. 1995) so in general
we adopt uncertainties of $\pm$0.5 mag in distance modulus/absolute magnitude 
for field and association  members.
Reduced uncertainties of $\pm$0.3 mag are restricted to
cluster members (NGC~3293, $h$ Per) and the more
spatially concentrated OB associations (Cyg OB7, Cyg OB9, Gem OB1, Cas OB14). 
For the two Ori OB1 members, we adopt distances  from individual Ori OB1 subgroups from
Brown et al. (1994), Ib for HD~37128 and Ic for HD~38771. The latter distance of 400pc 
is in preference to 220pc as derived from Hipparcos,  given the difficulty in determining 
individual distance estimates 
along this line of sight (de Zeeuw et al. 1999). For HD~190603, 
we adopt an absolute magnitude of --7.5$\pm$1.0 mag,
intermediate between normal Ia supergiants and the 
hypergiant HD~152236. Finally,
 we consider the Hipparcos membership of HD~53138 to Col~121 to be dubious
since it has insignificant proper motion compared to the cluster (de Zeeuw et al. 1999), and so instead
adopt an average B3\,Ia absolute magnitude. 

 Interstellar reddenings (assuming $R = A_{\rm V}/E(B-V)$ = 3.1 throughout)
are determined using intrinsic colours from Schmidt-Kaler (1982) and independently from
theoretical flux distributions determined in the present analysis, together with archival
large aperture $IUE$ spectrophotometry (except for HD~194279 and HD~148688).
Although the former was adopted for distance modulus/absolute magnitude estimates, the
latter provided a check on the reliability of intrinsic colours for early B supergiants.
In general the combined UV/optical flux distribution suggested intrinsic (B-V)$_{0}$ values 
$\sim$0.02 mag more negative than Schmidt-Kaler (1982) for our Ia sample.

Optical observations originate from either the Isaac Newton Group,
La Palma (northern sample) or CTIO, Chile (southern sample). 
The northern targets were originally observed in the blue visual
with the JKT (Lennon et al. 1992) with the Richardson-Brealey Spectrograph 
with the R1200B grating at a 
resolution of 0.8\AA\ and S/N$\sim$150. The majority were re-observed
with the ISIS spectrograph at the WHT at a similar resolution (Smartt,
priv. comm.) though higher S/N($\geq$300) 
using a setup identical to that used in Smartt et al. (2001). 
H$\alpha$ observations were obtained at the INT
using the IDS spectrograph by DJL, 
with a spectral resolution of 1.2\AA, and 
S/N$\sim$200. In three cases, HD~30614, HD~41117 
and HD~198478, WHT ISIS observations
were obtained by PAC during 2002 October
using a dichroic and 1.2arcsec wide slit, 
providing simultaneous blue and H$\alpha$ spectroscopy
with the  1200B/R gratings in first order, centred
at $\lambda$4400 and $\lambda$6650, with S/N$\geq$300.

The southern sample was observed at the CTIO 1.5m during May--June 2003 
via the SMARTS service program
using the Cassegrain spectrograph with Loral 1K (1200$\times$ 800 pixel)
CCD.
A standard reduction, bias subtraction, flat-fielding, and extraction
were carried out using IRAF. Blue and red spectroscopy of each
target was obtained using the 47 grating, providing a spectral range,
and (3 pixel) spectral resolution of 
660\AA, 1.6\AA\ (second order, blue) and 
1320\AA\, 3.0\AA\ (first order, red) 
respectively, as measured from HeAr and Ne arc lines. The extracted blue and
red continuum S/N ranges from 150 to 200, and 200 to 300 respectively.

In order to assess the robustness of our analyses based on
intermediate resolution observations covering a restricted wavelength
range, HD~115842 was observed for us by Dr C.J. Evans 
(ESO programme 073.D-0234) at the MPG/ESO 2.2m on 7 July 2004 with the 
Fibre-fed, Extended Range Echelle
Spectrograph (FEROS). A two arcsec aperture provides a resolving power
of R=48,000 covering a spectral range of 3500-9200\AA\ using an EEV
2kx4k CCD. A single exposure of 240s provided a continuum S/N of 300
at $\lambda$4600 in the pipeline reduced dataset (A. Kaufer, priv. comm.). 

Finally, (at least) 
two of the present sample definitely possess significant H$\alpha$ variability,
HD~37128 (B0\,Ia, Prinja et al. 2004) and  HD~14134 (B3\,Ia,  Morel et 
al. 2004). 
For HD~37128
we investigate the effect of variability on the derived mass-loss rates using H$\alpha$ observations
provided by Dr R.K. Prinja.

\begin{figure}[htbp]
\begin{center}
\includegraphics[width=0.7\columnwidth,clip,angle=-90]{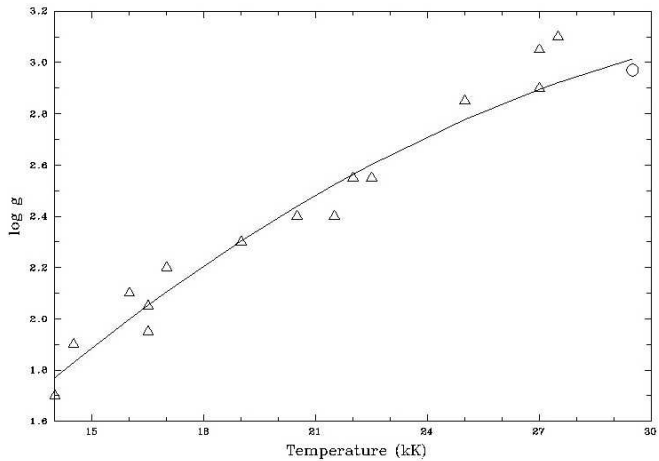}
\end{center}
\caption{$\log g-T_{\rm eff}$ calibration from Trundle 
et al. (2004) and Trundle \& Lennon (2005) for SMC B supergiants
(open triangles) plus HD~30614 from Repolust et al. (2005,
open circle).\label{logg}}
\end{figure}

\section{Analysis}\label{sect3}

\subsection{Methods}

Recent developments in model-atmosphere studies of hot stars routinely
permit sophisticated metal-line blanketing in non-LTE to be considered,
under either plane-parallel geometry (e.g., TLUSTY: 
Hubeny \& Lanz 1995;  Lanz \& Hubeny 2003)
or spherical geometry (e.g., CMFGEN: Hillier \& Miller 1998). 
The former technique provides a more comprehensive 
treatment of photospheric metal-line blanketing, while the latter offers 
insights into wind properties. 

In the present application, we have utilized CMFGEN, the current version
of which is discussed by Hillier et al. (2003).  The code solves the radiative
transfer equation in the co-moving frame, under the additional constraints of 
radiative and
statistical equilibrium.  The model used here is similar to that applied by 
Evans et  al. (2004b).  In total, 26 ions of H, He, C, N, O, Mg, Al, Si, S, Ca, 
Fe are included for early subtypes,  
comprising a total of 2,636 individual levels (grouped into
766 superlevels), with a full array of 25,960 bound-bound transitions. 
Higher ionization stages are omitted for B2.5--3 supergiants as indicated in
Table~\ref{table2}.

The 
temperature structure is determined by radiative equilibrium.  CMFGEN does not 
solve the momentum equation, so that a density or velocity structure is
required.  For the supersonic part, the velocity is parameterized with
a classical $\beta$-type law, with the exponent in the range $\beta$=1--3
selected on the basis of the H$\alpha$ profile. This is connected to a
hydrostatic density structure at depth, such that the velocity and velocity 
gradient match at the interface. The subsonic velocity structure is set by 
the corresponding H-He TLUSTY 
model which were calculated by the
authors for each CMFGEN model atmosphere calculation.
With regard to CMFGEN models in which the 
TLUSTY structure at depth is neglected, stellar temperatures are $\sim$1kK lower
and mass-loss rates 0.1--0.2 dex higher.

We initially estimated
surface gravities at individual temperatures from a 2nd order  fit to 
results from Trundle et al. (2004)
and Trundle \& Lennon (2005) for Ia B supergiants,  plus HD~30614 (O9.5\,Ia)
from  Repolust  et al. (2004). This is presented in Fig.~\ref{logg} and
should lead to gravities accurate to $\pm$0.2 dex. Subsequently, we have adjusted
surface gravities in TLUSTY models until H$\gamma$ is reproduced. We 
anticipate accuracies of 
0.1--0.15\,dex for all targets except perhaps the hypergiants HD~152236 and HD~190603 
which have significant wind contamination at H$\gamma$.

We have assumed a depth-independent Doppler profile for all lines when
solving for the atmospheric structure in the co-moving frame, while in the 
final calculation of the emergent spectrum in the observer's frame, we have 
adopted a radially dependent turbulence\footnote{In CMFGEN this 
adopted value is technically $\xi_{\rm min}$, the 
microturbulence at the base of the wind, which increases with radius to 
$\xi_{\rm max}$ = 50 \kms~at $v_\infty$.}, 
which reflects the effect of shocks due to wind instabilities.  For the 
present sample we initially assume $\xi$ = 20 \kms, with other values
in the range 10--40 \kms\ considered if consistent fits are not obtained 
for the helium and silicon lines.
 Incoherent 
electron scattering and Stark broadening for hydrogen and helium lines
are adopted.  We convolve our synthetic spectrum with a rotational broadening
profile.  Because of the intermediate dispersion of most observations, 
instrumental effects prevent a reliable determination of $v \sin i$
from the present dataset, such that we adopt the UV derived rotational 
rates from Howarth et al. (1997).

Terminal velocities, $v_\infty$, were uniformly determined from
$v_{\rm black}$ measurements from the literature (Howarth et al. 1997),
and are listed in Table \ref{results}. Alternatively, velocities may
be derived via the Sobolev with Exact Integral (SEI) technique (Lamers et al.
1987; Haser 1995).
Although individual measurements differ by as much as 20\%, 
$v_{\infty}$(Haser)/$v_{\infty}$(Howarth et al.)=1.02 for the 19 stars in
common to both samples. SEI modelling is often preferred to $v_{\rm black}$
measurements of wind velocity, although has its own modelling uncertainties. 
For example, Haser (1995) and Evans et al. (2004a) both used SEI modelling of
UV lines in HD~152236, deriving $v_{\infty}$=500 km\,s$^{-1}$ and
450 km\,s$^{-1}$, respectively, versus 390 km\,s$^{-1}$ adopted here from 
Howarth et al. (1997).

We varied the mass-loss rate and the 
velocity law (as characterized by the exponent $\beta$) until the H$\alpha$ 
profile was best reproduced in terms of intensity and morphology. Next, we
adjusted the stellar temperature to match the intensities of the
He\,{\sc ii} \lam4542 and He\,{\sc i} \lam4471 lines for O9.5--B0 stars,
Si~\4 \lam4089 and Si~\3 \lam4553-68-75 lines for B0.5--B2 stars, and
Si~\2 \lam4128-31 and Si~\3 \lam 4553-68-75 lines for B2.5--3 subtypes. 
The accuracy with which we can derive stellar temperatures is $\pm$1000K. 
Radii and luminosities follow from our derived
absolute magnitude, whilst mass-loss rates are obtained from a spectral
fit to the H$\alpha$ profile. With absolute magnitudes fixed at our adopted values,
formal uncertainties of 10\% in dM/dt follow for stars with
strong winds, increasing to 30\% for weak lined stars. In general,
the greatest uncertainty relates to the $\beta$ law, although the 
H$\alpha$ profile does provide some constraints on the $\beta$ exponent.

\begin{table*}
\caption{Derived non-LTE stellar and wind parameters. 
Two values are listed for the mass-loss rate  of HD~37128 since 
we fit the minimum and maximum H$\alpha$ profiles of Prinja et al. (2004).
 \label{table3}}
\begin{tabular}{
r@{\hspace{2mm}}l@{\hspace{2mm}}l@{\hspace{2mm}}l@{\hspace{3mm}}l@{\hspace{2mm}}l@{\hspace{0.5mm}}c
@{\hspace{2mm}}l@{\hspace{2mm}}r@{\hspace{2mm}}r@{\hspace{2mm}}r@{\hspace{0.5mm}}c@{\hspace{2mm}}c@{\hspace{2mm}}c
@{\hspace{2mm}}r@{\hspace{2mm}}c@{\hspace{2mm}}c}
\noalign{\hrule\smallskip}
HD & Name & Sp Type & $T_{\rm eff}$ & $\log  g$ & $\log g_{\rm eff}$ 
& $\log L/L_{\odot}$ & $R_{\ast}$  & $v_{\infty}$ & $v_{\infty}/v_{\rm esc}$ &
$\dot{M} \times 10^{6}$
 & $\beta$ &$v \sin i$ &  $\xi$ &  $M_{\rm V}$ & $\dot{M} v_{\infty} R^{0.5}$ \\ 
   &      &         & kK            &       cgs &  cgs &  & $R_{\odot}$ & \kms      & 
 & $M_{\odot}$/yr & \kms & \kms &\kms & mag & cgs\\ 
\noalign{\hrule\smallskip}
30614 & $\alpha$ Cam   & O9.5\,Ia& 29.0 & 3.0 & 2.75 & 5.63 &26.0& 1560 & 3.40 & 5.0 & 1.5 &129& 30 &  --6.6 & 29.40 \\ 
37128 & $\epsilon$ Ori & B0\,Ia   & 27.0  & 2.9 & 2.85& 5.44 &24.0& 1910 & 4.76 & (max) 2.5 &1.5 &91&12.5& --6.3 & 
29.17 \\ 
      &                &          &       &     &      &      &     &      
&    & (min) 2.0      &    &  &   &     &      29.07 \\
91969 &             & B0\,Ia   & 27.5  & 2.95 & 2.75   & 5.52  & 25.3 & 1470 & 3.33 & 1.0 & 1.5 &79&7.5 & --6.4 & 28.67\\ 
94909 &             & B0\,Ia   & 27.0  & 2.9 & 2.7    & 5.49 & 25.5 & 1050 & 2.55 & 2.0 & 1.5 &100$\ddag$ & 10 & --6.4 & 28.62\\ 
122879&                & B0\,Ia   & 28.0 & 2.95 & 2.7 & 5.52 &24.4& 1620 & 3.81 & 3.0 & 1.5&92&15  &  --6.4 & 29.18\\
38771 & $\kappa$ Ori & B0.5\,Ia   & 26.5  & 2.9 & 2.7 & 5.35 &22.2& 1525 & 3.84  & 0.9 &1.5&83&12.5& --6.1 & 28.61 \\ 
115842 &             & B0.5\,Ia   & 25.5 & 2.85 & 2.65 & 5.65 &34.2& 1180 & 3.86 & 2.0&1.5 &84&10& --6.9 & 28.94\\ 
152234 &         & B0.5\,Ia (N wk)&26.0& 2.85 & 2.6 & 5.87  & 42.4& 1450 & 2.85 & 2.7 & 1.5 &100$\ddag$& 10 & --7.4  & 
29.21\\
2905  & $\kappa$ Cas & BC0.7\,Ia& 21.5  & 2.6 & 2.45 & 5.52 &41.4& 1105 & 2.79 & 2.0 &2.0 &91&20&--7.1 & 28.95 \\ 
91943 &            & B0.7\,Ia     & 24.5  & 2.8 & 2.6 & 5.35 & 26.3 &1470 & 3.72  & 0.75 & 1.2&79&10& --6.3 & 28.55  \\
152235&            & B0.7\,Ia (N wk)&23.0 & 2.65 & 2.45 & 5.76  & 47.1 & 850 & 1.98 & 1.25  & 1.5 &81& 10 & --7.4 &  28.66 \\ 
154090&             & B0.7\,Ia    &  22.5 &2.65 & 2.45 & 5.48& 36.0&  915& 2.38 & 0.95&1.5 &78&10  & --6.8 & 28.52\\ 
13854 &             & B1\,Iab    & 21.5 & 2.55 & 2.4 & 5.43 &37.4&  920 & 2.67 & 0.85&2.0&97&20&--6.8 & 28.48\\
91316 & $\rho$ Leo & B1\,Iab (N str)&22.0  & 2.55 & 2.4 &5.47 & 37.4& 1110 & 3.31 &0.35&1.0&75&10& -6.8 & 28.17\\
148688 &           & B1\,Ia       &  22.0  & 2.6 & 2.4 &5.45 & 36.7&  725 & 1.99 &1.75&2.0&72&15& --6.8& 28.68 \\ 
14956  &  & B1.5\,Ia     & 21.0  & 2.5 & 2.35 & 5.65 & 50.6& 500: & 1.33: & 1.0 & 2.0 & 80$\ddag$ &10& --7.4 & 28.35 \\ 
152236 &       & B1.5\,Ia$^{+}$   & 18.0 & 2.2& 2.0 & 6.10 &112.4 &  390 & 1.01 & 6.0 &2.0&74&15&--8.8 & 29.19 \\ 
190603&        & B1.5\,Ia$^{+}$ & 18.5  & 2.25& 1.75& 5.57 &59.6 &  485 & 1.62 & 2.5 &3.0 &79&20&--7.5 & 28.77\\ 
14143 &        & B2\,Ia     & 18.0 & 2.25 & 2.05 & 5.42 & 52.9 & 645  & 2.21 & 1.05& 2.0&76& 20&--7.2 & 28.49 \\ 
14818 & 10 Per       & B2\,Ia & 18.5  & 2.4 & 2.25 & 5.35 &46.1&  565 & 1.66  & 0.55 &2.0 &82&20& --7.0& 28.12\\ 
41117 & $\chi^2$ Ori & B2\,Ia & 19.0  & 2.35  & 2.1 & 5.65 &61.9& 510 &  1.43   &  0.9  &2.0  &72& 10 & --7.6 &  28.36\\
194279 &       & B2\,Ia   & 19.0  & 2.3 & 2.1 & 5.37 &44.7& 550: & 1.99:& 1.05&2.5&80$\ddag$&20&  --7.0 & 28.39 \\ 
198478 & 55  Cyg     & B2.5\,Ia & 16.5 & 2.15 & 1.75 & 5.03 &40.0&  470 & 2.02 & 0.23 &2.0&61&20&--6.4 & 27.49\\ 
14134 &              & B3\,Ia   & 16.0 & 2.05&1.75 & 5.28 &  56.7 &  465 & 1.94 & 0.52 & 2.0 &66& 15 &--7.1 & 28.06 \\ 
53138 &$o^2$ CMa     & B3\,Ia   & 15.5 & 2.05& 1.75 & 5.34  & 65.0  &  865 & 3.25 & 0.36 &2.0&58&20&--7.3 & 28.20 \\  
\noalign{\hrule\smallskip}
\end{tabular}
$\ddag$: Our study requires higher $v \sin i$ values for HD~94909 and 
HD~152234 than 68 and 76 km\,s$^{-1}$ obtained by Howarth et al. (1997),
respectively, whilst HD~194279 and HD~14956 were not included in their sample.
\end{table*}

He/H abundances are relatively difficult to constrain in B supergiants
 since different choices of He\,{\sc i} lines yield different He 
abundances, in part due to the well known singlet-triplet problem 
(Lennon et al. 1991; McErlean et al. 1998). 
Consequently, we adopt a uniform He/H = 0.2 by number for all our targets. 
This was selected on the basis of the average N/C enrichment derived relative to 
Solar values (Sect.~\ref{abundances}), with respect to recent interior 
evolutionary models for rotating massive stars (Meynet \& Maeder 2000).
Fortunately, an error in He/H will only weakly affect the derived stellar temperature
and/or mass-loss rates.  Indeed, Repolust et al. (2005) have recently obtained a
typical He/H abundance of 0.2 amongst their sample of Galactic OB supergiants.

CNO abundances were varied to fit the
relevant optical lines, with a typical accuracy of $\sim$0.2 dex.     
Solar abundances were taken from Grevesse \& Sauval (1998) except
oxygen, which was set at log(O/H)+12 = 8.66 (Asplund et al. 2004).  
Abundances of other metallic elements were fixed at Solar
values. The primary diagnostic lines used for determination of the nitrogen 
abundance were N~\2 \lam3995 and \lam4601-43, plus 
 N~\3 \lam4097 blend for the earliest subtypes where N~\2 is weak/absent.
Oxygen and carbon abundances were also determined from consideration of 
the  optical lines; O~\2  \lam4069--4092 and \lam4590-96 provided useful  
constraints on the oxygen abundance, whilst C~\2 \lam4267 together
with C~\3 \lam4647--51 at early subtypes (blended with O~\2) 
were used for carbon. We anticipate accuracies in derived 
CNO abundances to be no greater than $\sim$0.15 dex (O) to 
$\sim$0.3 dex (C and N), where differences are due to the number of diagnostics
available.

\begin{figure*}[ht!]
\begin{center}
\includegraphics[width=1.75\columnwidth,clip,angle=0]{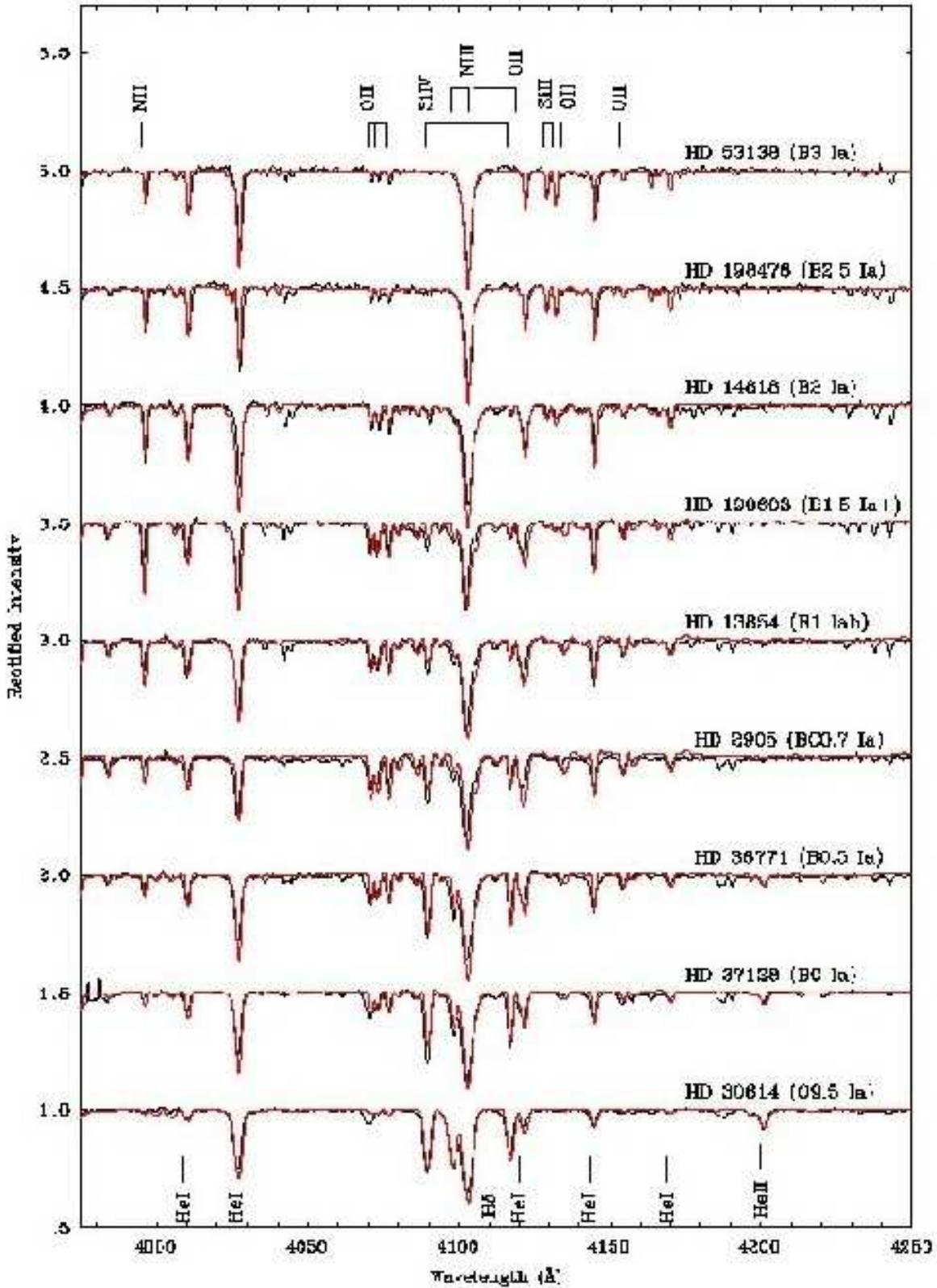}
\end{center}
\caption{(a) Synthetic fit (red in electronic version, dotted in paper version) 
to $\lambda\lambda$3975--4250 spectral range for representative early B supergiants (solid black).\label{fig1a}}
\end{figure*}

\addtocounter{figure}{-1}

\begin{figure*}[ht!]
\begin{center}
\includegraphics[width=1.75\columnwidth,clip,angle=0]{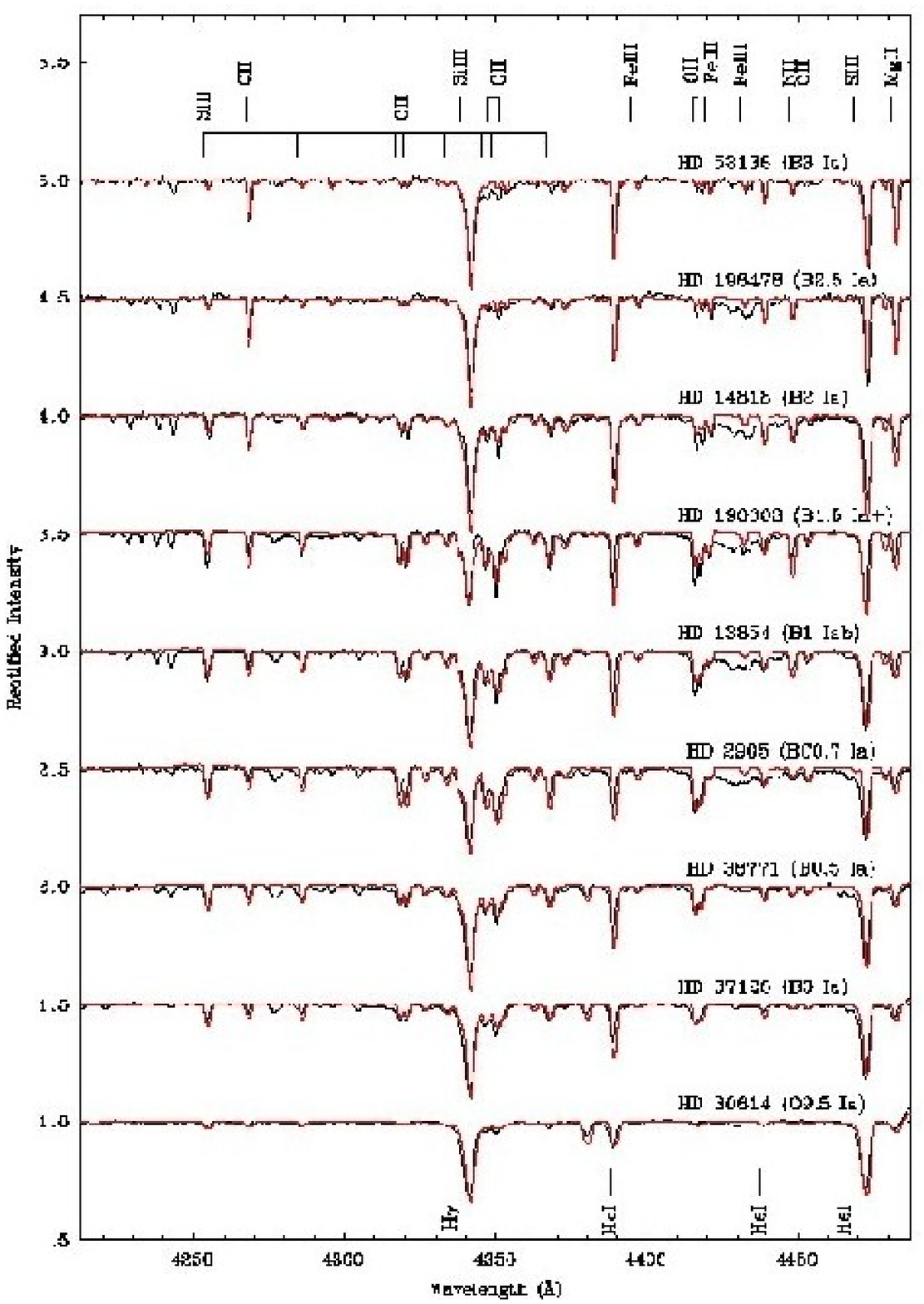}
\end{center}
\caption{(b) Synthetic fit to $\lambda\lambda$4215--4485 spectral 
range for representative
early B supergiants.\label{fig1b}}
\end{figure*}

\addtocounter{figure}{-1}

\begin{figure*}[ht!]
\begin{center}
\includegraphics[width=1.75\columnwidth,clip,angle=0]{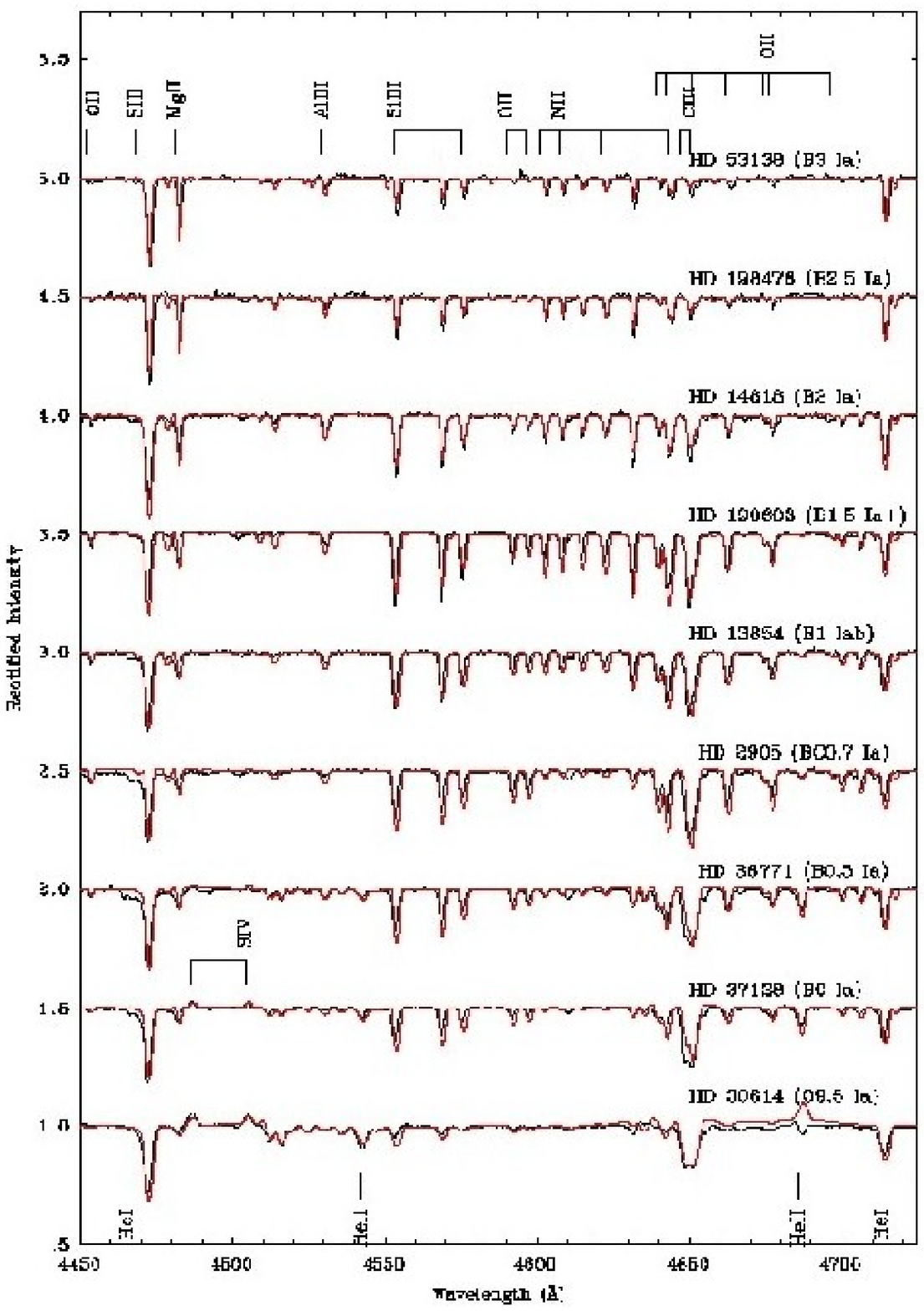}
\end{center}
\caption{(c) Synthetic fit to $\lambda\lambda$4450--4725 
spectral  range for representative early B supergiants.\label{fig1c}}
\end{figure*}

\begin{figure*}[ht!]
\begin{center}
\includegraphics[width=1.4\columnwidth,clip,angle=-90]{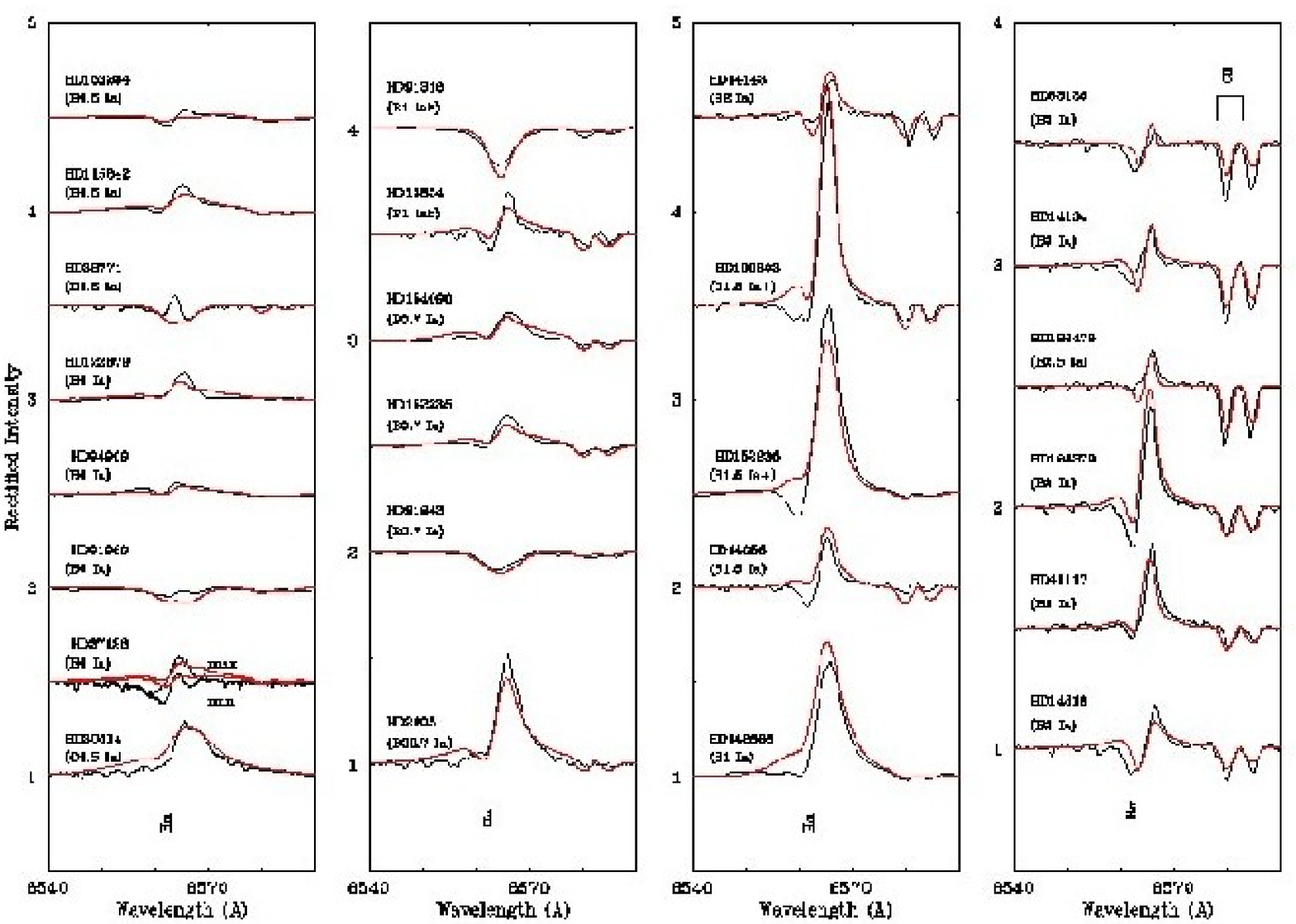}
\end{center}
\caption{Spectral fits (red in electronic version, dotted in paper version) 
 to H$\alpha$ and C\,{\sc ii} $\lambda$6578--82 
for O9.5--B3 supergiants.\label{fig3}}
\end{figure*}

\section{Spectroscopic Results}\label{results}

Derived stellar, wind and chemical properties of our sample of OB supergiants
based on the intermediate dispersion spectroscopy are provided in Table~\ref{table3}. 
We include surface and effective gravities, corrected for the Eddington parameter, $\Gamma_e$,
which is the ratio of the radiative acceleration from electron scattering 
to the stellar gravity.

\subsection{Physical and wind properties}

Figure~\ref{fig1a}(a-c) compares blue optical spectroscopy for selected 
O9.5--B3\,Ia supergiants with synthetic spectral fits. In general, the quality
of spectroscopic fits is excellent, not only for our selected He or Si
diagnostic lines,
but also for classification lines such as He\,{\sc i} $\lambda$4471 vs Mg\,{\sc ii}
$\lambda$4481,  plus weak features of other non-CNO elements, e.g. Al\,{\sc iii}
$\lambda$4529 and S\,{\sc iv} $\lambda$4485--4504. The major differences reflect
weak metal transitions, which are missing from our selected model atoms, with no
effect upon the derived physical properties. Apparent
discrepancies in the region of $\lambda$4430 are in fact due to the broad
Diffuse Interstellar Band (DIB) at that wavelength, whose strength scales with
interstellar reddening. The only clear discrepancies occurs for HD~30614 (O9.5\,Ia)
and HD~37128 (B0\,Ia) at He\,{\sc ii} $\lambda$4686 and Si\,{\sc iv} $\lambda$4088-4116.

In Fig.~\ref{fig3} we present spectral fits to all our program stars in the 
vicinity of H$\alpha$, with model profiles degraded to the spectral resolution
of the individual observations. Overall, agreement is excellent, except that the blue 
P~Cygni absorption component is imperfectly reproduced for the B1--2 supergiants.
Note that the C\,{\sc ii} $\lambda\lambda$6578-82 doublet
is well matched in mid B supergiants in most cases. As already indicated,
HD~37128 (B0\,Ia) is observed to exhibit large scale H$\alpha$ variability
(Prinja et al. 2004) so we have attempted to reproduce the minimum and
maximum emission cases to quantify the range in derived mass-loss 
rate, namely 2.0--2.5 $\times 10^{-6}$ $M_{\odot}$ yr$^{-1}$.

\begin{figure}[htbp]
\begin{center}
\includegraphics[width=0.7\columnwidth,clip,angle=-90]{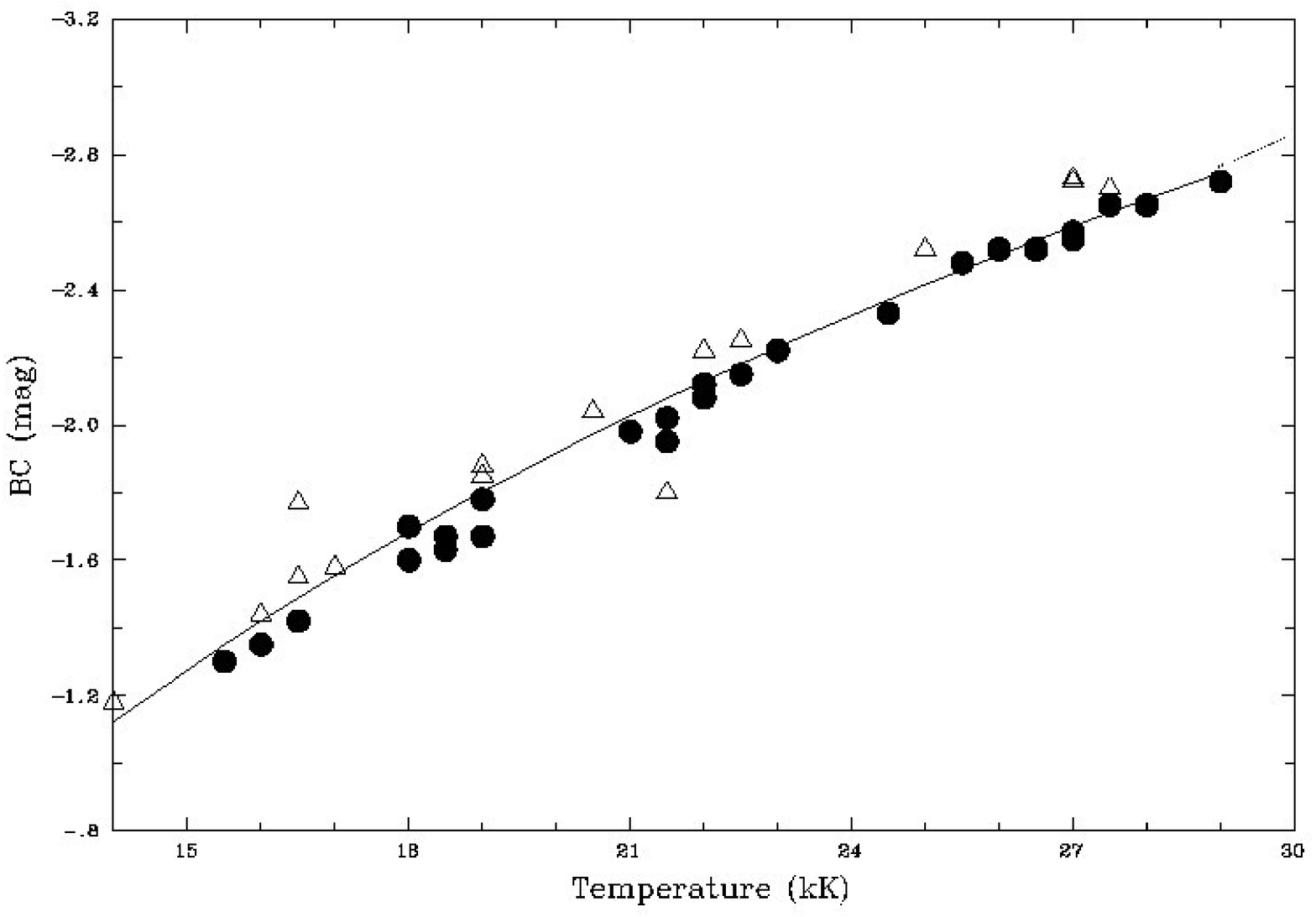}
\end{center}
\caption{Bolometric correction (BC) as a function of temperature for 
Galactic (filled circles, this study) and SMC (open triangles, Trundle 
et al. 2004; Trundle \& Lennon 2005) B supergiants. The
fit to the combined sample (solid line) connects well to the
recent Martins et al. (2005) relation for O stars (dotted line). 
\label{bol_cor}}
\end{figure}

Table~\ref{bstars} compares our derived temperatures for B supergiants 
with those derived previously
by McErlean et al. (1999), using hydrostatic, plane-parallel model atmospheres. As with O stars (e.g.
Crowther et al. 2002, Repoulst et al. 2004), current line-blanketed, spherically extended 
model atmospheres indicate a downward revision in the temperature scale, by typically 1--2kK 
Indeed, stellar temperatures obtained with CMFGEN seem to be agree well with FASTWIND (Puls et al.
2005), as indicated
by Urbaneja (2004) who has used FASTWIND to study HD~37128, HD~38771, and HD~14956 
revealing temperatures consistent with our results to $\sim$0.5kK. Given this close agreement,
we compare bolometric corrections (BCs) as a function of temperature 
from the present sample, supplemented by SMC B  supergiant results
from Trundle et al. (2004) and Trundle \& Lennon (2005) in Fig.\ref{bol_cor}. A fit to the
combined dataset indicates
\[ {\rm BC} = 20.15 - 5.13 \log T_{\rm eff} \]
with a standard deviation of $\sim$0.05 mag, which 
closely matches the recent result for O stars by Martins et al. 
(2005) at $\sim$30kK.

Spectroscopically estimated 
masses span a wide range, from
$8^{+4}_{-2} M_{\odot}$ (HD~198478) to 76$^{+44}_{-22} M_{\odot}$ 
(HD~152236), based on gravities accurate to $\log g \pm$0.15 dex. In general,
spectroscopic masses are lower than estimates of the evolutionary masses, indicating
another case of a `mass discrepancy', as already
established by  Trundle et al. (2004) for SMC B supergiants. For example, we estimate
an initial (current) evolutionary mass of 20$M_{\odot}$ (18 $M_{\odot}$) 
for HD~198478. In order to resolve the spectroscopic and evolutionary masses, we would 
require $\log g\sim $2.5, versus 2.15$\pm$0.15 obtained for this star.

In Table~\ref{table4} we compare the present 
results with those from previous studies, namely
(i) Barlow \& Cohen (1977) based on the mid-IR free-free excess technique of 
Wright \& Barlow (1975)
and adopted temperature scale for B supergiants, and (ii)  Kudritzki et al. (1999) who
adopted the temperature scale of McErlean et al. (1999), based on hydrostatic, plane-parallel
model atmospheres and derived  H$\alpha$ mass-loss rates with the unblanketed version
of FASTWIND (Santolaya-Rey et al. 1997); 
(iii) Urbaneja (2004) based on FASTWIND (Puls et al. 2005) model atmospheres.

\begin{table}[htbp]
\caption{Comparison between our derived non-LTE line blanketed 
temperature scale (in kK) of Galactic  early B\,Ia and Iab
supergiants and 
the unblanketed temperature scale of McErlean et al. (1999), 
plus recent line blanketed Magellanic Cloud results from Trundle et al.
(2004), Trundle \& Lennon (2005). For each study the range in temperatures
and number of stars studied is indicated. 
Values for hypergiants are flagged with $\ast$,
and are excluded from our current ``mean'' temperature scale for Ia 
supergiants.}
\label{bstars}
\begin{tabular}{l@{\hspace{2mm}}c@{\hspace{0.5mm}}c@{\hspace{2mm}}c@{\hspace{0.5mm}}c
@{\hspace{2mm}}c
@{\hspace{0.5mm}}c
@{\hspace{1mm}}r}
\noalign{\hrule\smallskip}
Subtype &  \multicolumn{2}{c}{McErlean et al.}& \multicolumn{2}{c}{Trundle et al.} &\multicolumn{2}{c}{This work}&$T_{\rm eff}$ \\ 
 &  \multicolumn{2}{c}{Galaxy/unblank} &\multicolumn{2}{c}{SMC/blank}         & \multicolumn{2}{c}{Galaxy/blank} & mean\\
\noalign{\hrule\smallskip}
O9.5\,Ia &  32.5    & 1 &    &  & 29.0 & 1  & 29\\
O9.7\,Ia &  29      & 1 &    &  &      &    & 28.5 \\
B0\,Ia   &  28.5    & 2 & 27 & 1& 27--28 & 4 & 27.5\\
B0.5\,Ia &  27.5    & 1 & 27 & 1& 25.5--26.5 & 3 & 26\\
B0.7\,Ia &  24      & 1 &    &  & 21.5--24.5 & 4 & 22.5\\
B1\,Ia   &  23.5    & 1 &21.5$^{\ast}$--25 & 3&21.5--22 & 3 & 21.5\\
B1.5\,Ia &  21$^{\ast}$--21.5 & 2 & 20.5-22.5 & 3& 18$^{\ast}$--21.5 & 
3 & 20.5\\
B2\,Ia   &  19-20   & 4 & 19 & 2 & 18--19 & 4 & 18.5\\
B2.5\,Ia &  18      & 1 & 16--17 & 4& 16.5 & 1& 16.5\\
B3\,Ia   &  17--18.5 & 4 & 14 & 1 & 15.5--16 & 2 & 15.5\\
\noalign{\hrule\smallskip}
\end{tabular}
\end{table}

\begin{table}[htbp]
\caption{Comparison between present results for selected B supergiants 
and those determined from (a) mid-IR free-free excess method of Barlow \& Cohen (1977, BC77),
and (b) Kudritzki et al. (1999, K99) following the H$\alpha$
fitting technique of Santolaya-Rey et al. (1997) plus the plane-parallel 
hydrostatically derived temperature scale from McErlean et al. (1999).
For HD~30614 we compare the present results with both Puls et al. (1996, P96) and the recent line blanketed
study of Repolust et al. (2004, R04), whilst we also include line blanketed results from Urbaneja (2004, U04)
for HD~37128 and HD~38771. Wind velocities in parenthesis relate to adopted values.\label{table4}}
\begin{tabular}{r@{\hspace{2mm}}l@{\hspace{1mm}}c@{\hspace{1mm}}c@{\hspace{1mm}}c@{\hspace{1mm}}l@{\hspace{1mm}}
l@{\hspace{1mm}}l}
\noalign{\hrule\smallskip}
HD & Sp Type & $T_{\rm eff}$ & $\log (L/L_{\odot})$ & $v_{\infty}$ & $\dot{M} \times 10^{6}$ & M$_{\rm V}$ & Ref \\ 
   &         & kK            &                      &  km/s        & $M_{\odot}$yr$^{-1}$ & mag \\
\noalign{\hrule\smallskip}
30614 & O9.5\,Ia & 30.0       &   5.79               &  1550        & 5.2        & --7.0   & P96\\
      &        & 29.0        & 5.83                 & 1550         & 6.0        & --7.0  & R04 \\
      &        & 29.0        & 5.63                 & 1560         & 5.0        & --6.6  & This work \\
37128 & B0\,Ia &  25.0         & 5.62                 & 2300       & 1.7 & --6.8 & BC77 \\
      &      & 28.5          & 5.86                 & 1600       & 2.4 & --7.0 & K99\\
      &      & 27.5          & 5.69                 & 1600       & 2.5 & --6.9 & U04 \\
      &      & 27.0          & 5.44                 & 1910        & 2--2.5       & --6.3 & This work\\
38771 & B0.5\,Ia& 22.0         & 5.30                 & 1900        & 0.7       & --6.3 & BC77 \\ 
      &       & 27.5         & 4.94                 & 1350         & 0.27      & --4.8 & K99\\
      &       & 26.5         & 4.94                 & 1350         & 0.48      & --4.9 & U04 \\
      &       & 26.5       & 5.35                 & 1525         & 0.9   & --6.1 & This work\\
2905 & BC0.7\,Ia& 20.0         & 5.46                 &(1500)        & 1.4       & --6.9 & BC77 \\
     &        & 24.0           & 5.70                 & 1100         & 2.3       & --7.0 & K99\\
     &        & 21.5         & 5.52                 & 1105         & 2.0       & --7.1 & This work\\
13854 & B1\,Iab &23.5          & 5.53                 & 1000.       & 0.78       & --6.7 & K99\\
      &       & 21.5         & 5.43                & 920         & 0.85        & --6.8 & This work\\
91316 & B1\,Iab & 21.0         & 5.22                 & 1580      & 0.86         & --6.4 & BC77 \\
     &        & 22.0         & 5.47                 & 1110       & 0.35         &  --6.8 &This work\\
14956 & B1.5\,Ia & 20.5        & 5.66                 & (1200)       & 1.2          & --7.0 & U04 \\
      &        & 21.0        & 5.65                 & (500)      & 1.0          & --7.4 & This work \\
190603& B1.5\,Ia$^{+}$& 19.0   & 5.66                 & (1000)      & 2.9        & --7.5 & BC77 \\
      &       &   19.5      &  5.63                    &  485     & 2.5        & --7.5 &This work \\ 
14143 & B2\,Ia  & 18.0         & 5.46                 & (1000)      & 1.2        & --7.1 &BC77\\ 
      &       & 20.0        & 5.51                  & 650.        & 0.30       & --6.95& K99 \\
      &       & 18.0        & 5.42                  & 645         & 1.05      & --7.2 & This work \\
14818 & B2\,Ia  & 20.0        & 5.47                  & 650.        & 0.25       & --6.9 & K99\\
      &       & 18.5        & 5.35                  & 565         & 0.55        & --7.0 & This work\\
41117 & B2\,Ia & 18.0        & 5.58                   & (1000)       & 1.2      & --7.4 & BC77 \\
      &        & 19.5        & 5.70                   & 500          & 0.85     & --7.5 & K99 \\
      &        & 19.0       &  5.65                      & 510       & 0.9       & --7.6  & This work \\
198478& B2.5\,Ia& 15.0     & 5.16                    & (580)    & 0.56    & --6.8 & BC77 \\
      &        & 16.5      & 5.03                    & 470      & 0.23    & --6.4 & This work \\
14134 & B3\,Ia  & 15.0        & 5.24                   & (580)        & 0.71      & --7.0 & BC77 \\ 
      &       & 18.0        & 5.48                  & 465         & 0.15       & --7.1 & K99\\
      &       & 16.0       &  5.28                  & 465            &  0.52    & --7.1 & This work \\
53138 & B3\,Ia  & 15.0        & 5.28                  & 580         & 1.7        & --7.1 & BC77 \\ 
      &       & 18.5        & 5.22                  & 620         & 0.095      & --6.4 & K99\\
      &       & 15.5        & 5.34                  & 865         & 0.36        & --7.3 & This work\\
\noalign{\hrule\smallskip}
\end{tabular}
\end{table}

Kudritzki et al. (1999) adopted the Hipparcos distance to HD~38771, whilst we adhere its 
the membership of Ori~OB1c. Putting HD~38771 and the H$\alpha$ variable HD~37128 aside,
consistency between Kudritzki et al. and the present sample
is excellent for the other B0--1 supergiants in common, 
with $\dot{M}_{{\rm H}\alpha}/\dot{M_{\rm K99}} \sim$ 1$\pm$0.1. 
In contrast, the present results are systematically higher for B2--3 supergiants, 
with $\dot{M}_{{\rm H}\alpha}/\dot{M_{\rm K99}} \sim$ 3, except for HD~53138 for which
substantially different absolute magnitudes were adopted in the two studies. 
This result, together with the good consistency between our results using CMFGEN and FASTWIND 
(Urbaneja 2004) suggests problems with the application of the analytical mass-loss technique 
(originally devised for O stars  by Puls et al. 1996) to B2--3 supergiants by Kudritzki et al. 
(1999).

\begin{figure}[htbp]
\begin{center}
\includegraphics[width=0.7\columnwidth,clip,angle=-90]{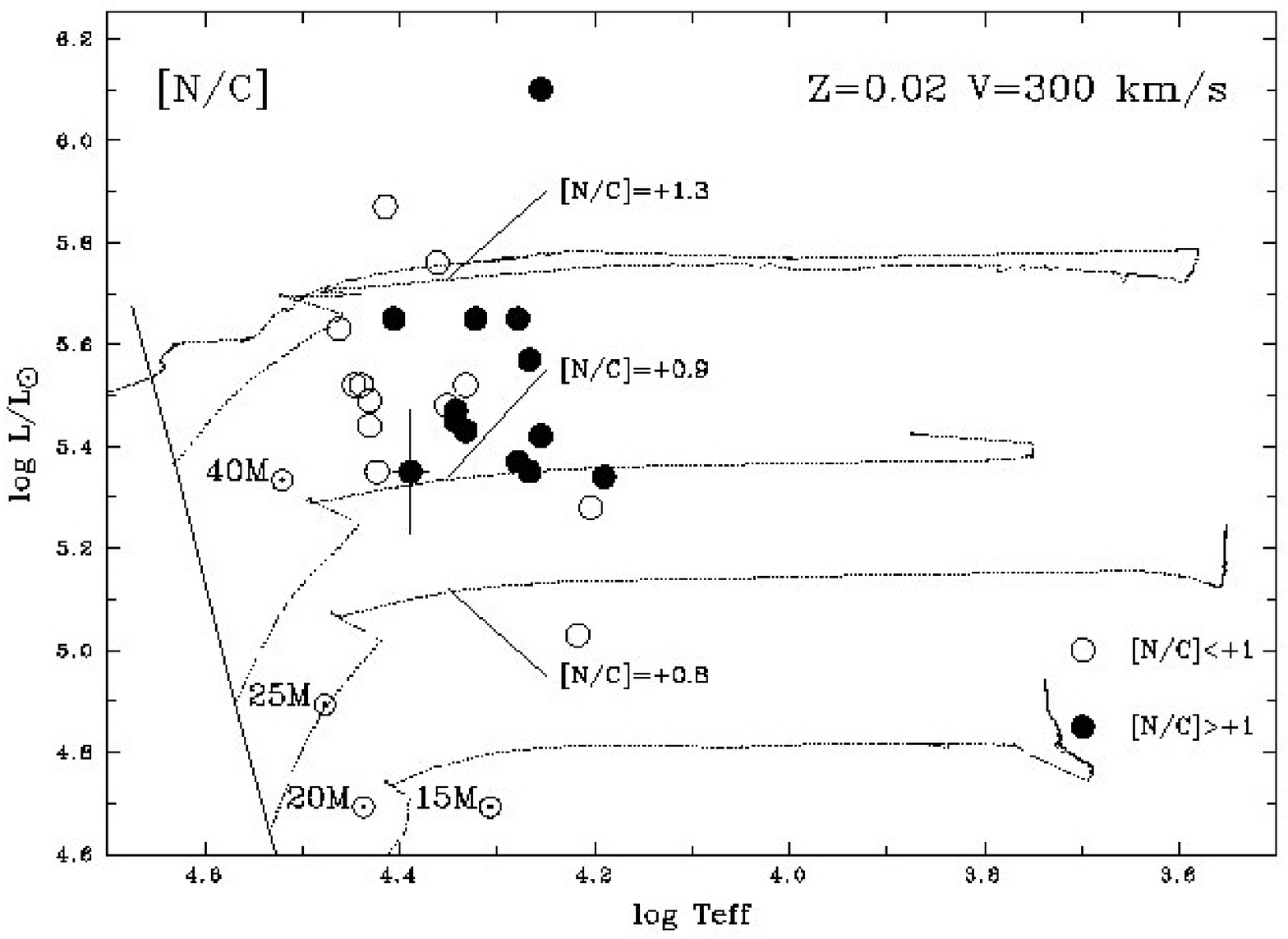}
\end{center}
\begin{center}
\includegraphics[width=0.7\columnwidth,clip,angle=-90]{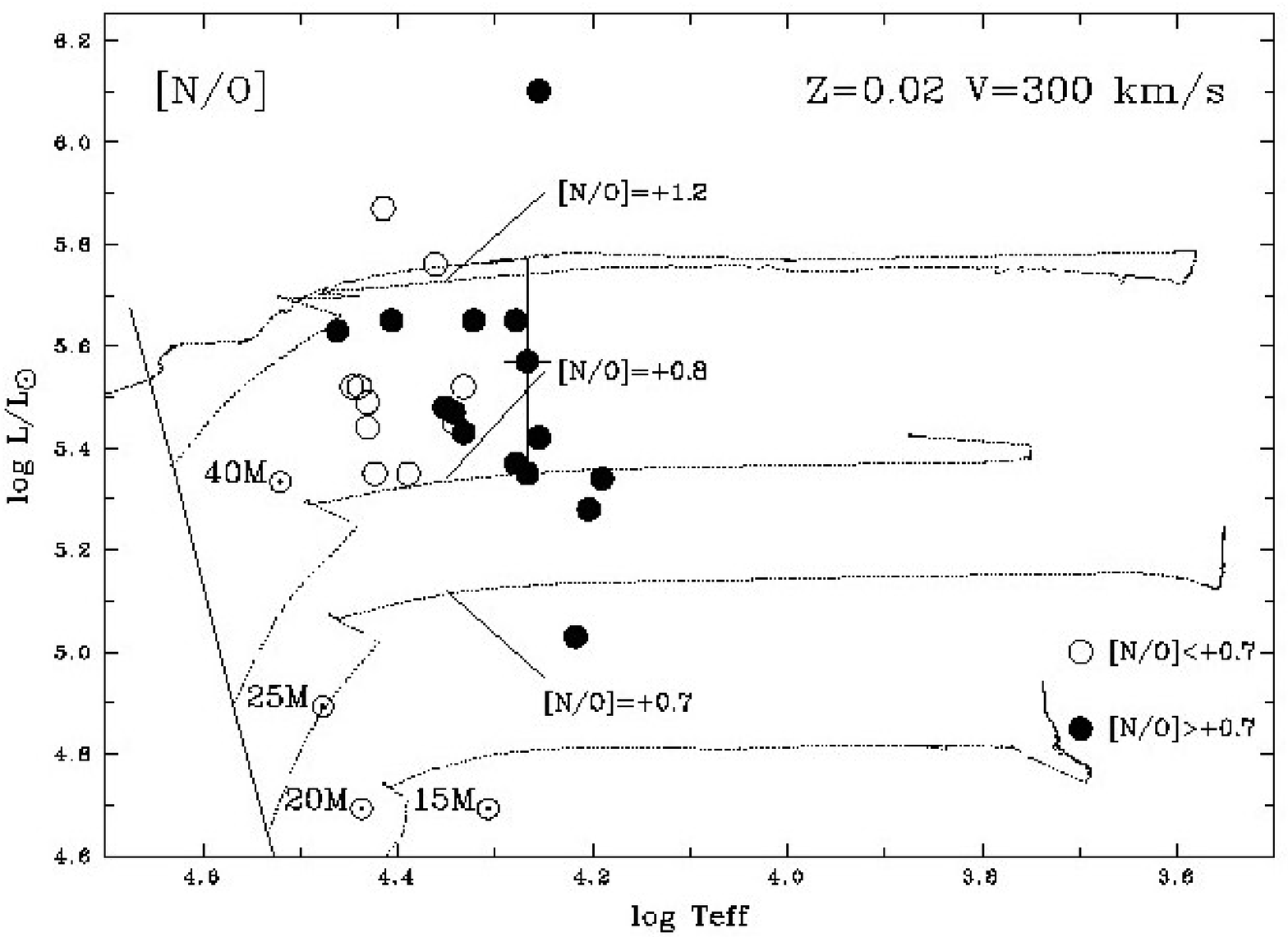}
\end{center}
\caption{{\bf (Upper panel)} Observed H-R diagram for program stars, coded by degree of N/C enrichment,
together with tracks from Solar metallicity, rotating (300 km/s) evolutionary models (dotted lines)
from Meynet \& Maeder (2000), for which predicted N/C enrichments are indicated at $T_{\rm eff}$=22,000K.
The error bar for HD~91943 is indicated ($\pm$0.3 mag uncertainty in distance modulus);
{\bf (Lower panel)} As above for N/O. The error bar for HD~190603 is indicated 
($\pm$1.0 mag uncertainty in distance modulus).}
\label{hrd}
\end{figure}

For those 10 stars in common with Barlow \& Cohen (1977), if we adjust
their results to our adopted wind velocities and distances, we find ratios
$\dot{M}_{{\rm H}\alpha}/\dot{M}_{\rm mid-IR}$ in the range 0.7--2.3 with an average of 
1.4, excluding HD~53138 for which the mid-IR continuum exceeds the H$\alpha$ result
tenfold. Kudritzki et al. (1999) also comment on the peculiar mid-IR brightness
of HD~53138 with regard to optical and radio mass-loss indices. 
Overall, agreement is satisfactory between the two techniques. 
This is of particular interest, given clumping, if present, is predicted to 
affect these measurements in different ways -- H$\alpha$ samples the highest density 
region immediately above the photosphere, whilst the mid-IR continuum samples free-free emission ,
superimposed upon the stellar photosphere at larger radii (Runacres \& Owocki 2002).
Spectroscopic studies in which clumping is considered generally produce superior spectral
fits to far-UV lines (e.g. Evans et al. 2004b). Consequently, a 
comparison with Barlow \& Cohen (1977) suggests comparable clumping factors
in the H$\alpha$ and mid-IR continuum forming regions.

\subsection{Elemental abundances}\label{abundances}

Elemental abundances are presented in Table~\ref{cno}. These 
represent the first attempt
to quantify metal abundances in a substantial sample of Galactic early 
and mid
B supergiants. With respect to current Solar values (Asplund et al. 2005), carbon
is depleted by 0.0--0.8 dex, nitrogen is enriched by 0.1--1 dex, with oxygen typically
mildly depleted (up to 0.5dex). 
Consequently, the mean [N/C] and [N/O] ratios for our
sample, +1.07 and +0.74 dex respectively, 
indicate evidence for partial 
(rotationally mixed?) CNO processing at their surfaces. 

\begin{table}[htbp]
\caption{Individual and mean CNO abundances for Galactic B supergiants (given as
$\log N(X)/N(H)+12$), where $X$ represents the appropriate element, versus Solar abundances
(Asplund et al. 2005), to the nearest 0.05 dex. For all stars He/H=0.2 by number was adopted.
In addition, we show the mean non-LTE abundances obtained from two  B1-2Ib
supergiants from Gies \& Lambert (1992, GL92), four  B2Ia supergiant results from Lennon (1994, L94), 
plus luminosity class II-Ia Galactic A supergiants from Venn (1995, V95).
\label{cno}}
\begin{tabular}{l@{\hspace{2mm}}l@{\hspace{2mm}}l@{\hspace{2mm}}c@{\hspace{2mm}}c
@{\hspace{2mm}}c
@{\hspace{2mm}}l
@{\hspace{2mm}}l}
\noalign{\hrule\smallskip}
HD & Sp Type & $T_{\rm eff}$/kK & C & N & O & [N/C] & [N/O]\\
Solar &          &       & 8.39 & 7.78   & 8.66 & 0.0 & 0.0 \\ 
\noalign{\hrule\smallskip}
30614 & O9.5\,Ia& 29.0 & 8.15 & 8.3  & 8.35 & +0.8 & +0.85 \\ 
37128 & B0\,Ia   & 27.0 & 7.95&8.15& 8.55     & +0.8 & +0.5 \\ 
91969 &  B0\,Ia &  27.5&   8.15 & 8.15 & 8.65  & +0.65     & +0.4    \\ 
94909  & B0\,Ia   & 27.0  & 7.95  & 8.15  & 8.65 & +0.8 & +0.4   \\ 
122879 & B0\,Ia   & 28.0  &7.95&8.15& 8.65 & +0.8 & +0.4 \\ 
38771  & B0.5\,Ia & 26.5 &7.95&8.25&8.55 & +0.9 & +0.6\\ 
115842 & B0.5\,Ia  & 25.5 & 7.80&8.25&8.45 & +1.1 & +0.7\\ 
152234 & B0.5\,Ia (N wk)&26.0& 7.95 &  8.05 & 8.75 &+0.7 & +0.2   \\ 
2905  & BC0.7\,Ia& 21.5  & 7.9 &8.0&8.75 & +0.7 & +0.1 \\ 
91943 & B0.7\,Ia  & 24.5 & 7.65 & 8.15& 8.4 & +1.1 & +0.6  \\ 
152235& B0.7\,Ia (N wk)&23.0 & 7.95 & 8.0: &8.55 & +0.65 & +0.3  \\ 
154090& B0.7\,Ia    &  22.5 &7.95&8.3&8.4 & +1.0 & +0.8\\ 
13854 & B1\,Iab    & 21.5  &7.95&8.45&8.55 & +1.1 & +0.8\\ 
91316 & B1\,Iab (N str)&22.0  & 7.5&8.3 & 8.4 & +1.45 & +0.8\\  
148688 &            B1\,Ia &  22.0  &7.65&8.15&8.55 &+1.1 & +0.5\\ 
14956  &  B1.5\,Ia  & 21.0 & 7.95&8.75 & 8.15  & +1.4 & +1.5\\  
152236 & B1.5\,Ia$^{+}$   & 18.0   &7.30&8.75&8.3 & +2.05 & +1.3\\ 
190603        & B1.5\,Ia$^{+}$ & 18.5  &7.95&8.75&8.75 & +1.4 & +0.9 \\ 
14143               & B2\,Ia     & 18.0 &7.6   & 8.7  & 8.6 & +1.75 & +1.0    \\ 
14818 & B2\,Ia & 18.5  &7.65&8.35&8.45 & +1.3 & +0.8 \\ 
41117 & B2\,Ia & 19.0 & 7.65   & 8.55  & 8.45  & +1.5  & +1.0 \\ 
194279        & B2\,Ia   & 19.0&7.95&8.65&8.45 & +1.3 & +1.1 \\ 
%
198478 & B2.5\,Ia & 16.5  &8.25&8.25&8.3 &+1.0 & +1.2\\ 
14134 & B3\,Ia   & 16.0 &8.25&8.45&8.45    & +0.8  & +0.9 \\ 
53138 & B3\,Ia   & 15.5  &7.95&8.45&8.15 & +1.1 & +1.2 \\ 
\noalign{\hrule\smallskip}
Mean  &          &       & 7.93 &  8.42  &  8.51 & +1.10  & +0.79  \\
\noalign{\hrule\smallskip}
GL92   & B1-2\,Iab/b     &         & 7.92 & 8.30 &  &  +0.99    \\
L94    & B2\,Ia       &         &      &      &    & +1.81    & +0.65\\ 
V95   & A0-F0       &     & 8.14 & 8.05 &      &+0.52 &    \\
\noalign{\hrule\smallskip}
\end{tabular}
\end{table}

These conclusions support the relatively few quantitative abundance estimates
for B supergiants from the literature. Previous studies 
typically relied upon differential abundance analyses (e.g. Smartt et al. 1997) 
using blanketed LTE model atmospheres (Kurucz 1991), supplemented by non-LTE plane parallel,
hydrostatic model atmospheres (e.g. Lennon et al. 1991, Gies \& Lambert 1992). We
include results for four morphologically normal B2\,Ia supergiants (including HD~14818
and HD~14143) from Lennon (1994), plus two B1--2\,Iab/b supergiants (including HD~91316)
from Gies \& Lambert (1992) in Table~\ref{cno}. For stars in common between the studies, 
Gies \& Lambert (1992)
obtained [N/C]=+1.11 dex (we derive +1.45) for HD~91316 (B1\,Iab) and [N/C]=+1.13 dex 
(versus +0.98 dex  here) for HD~198478 (B3\,Ia), 
whilst Lennon (1994) obtained [N/C]=+1.6 dex (we derive +1.28 dex) and [N/O]=+0.58 (versus
+0.78 dex) for HD~14818 (B2\,Ia), and similar differences for HD~14143 (B2\,Ia). Venn (1995)
obtained rather lower nitrogen enrichment in a sample of typically lower mass  A supergiants
and bright giants (see Table~\ref{cno}).

In Fig.~\ref{hrd} we present the observed H-R diagram for our program stars for [N/C] and
[N/O] coded by the degree of enrichment, together with initially rotating  (300 km/s)
evolutionary models for Solar metallicity from Meynet \& Maeder (2000). In general, greater
nitrogen enrichment is observed in lower temperature stars, whilst greater enrichment is
predicted at higher luminosity. Note that the lower luminosity mid-B supergiants 
appear to show greater processing of oxygen than carbon from Fig.~\ref{hrd}.

For each of the 20, 25 and 40$M_{\odot}$ tracks we present the predicted enrichment at a temperature
of 22kK. On average, the predicted [N/C] and [N/O] ratios of +0.8--1.3 and +0.7--1.2
match observations rather well, with the highest luminosity star HD~152236 revealing the
greatest nitrogen enrichment, also as predicted.

Within our sample, all are morphologically normal B supergiants, except that 
$\kappa$ Cas has a BC subtype (Walborn 1972), suggesting a lower degree of CNO processing,
In fact, HD~2905 does have the least enriched N abundance of our sample (two
N wk supergiants are also only moderately enriched), but does appear to be 
moderately depleted in C, with no significant processing in O. Consequently,
HD~2905 does indeed possess the lowest [N/O] and [N/C] of our sample, confirming its
morphological difference, potentially due to a low initial rotation velocity.

Finally, Urbaneja (2004) has used FASTWIND to
derive physical properties and metal abundances of several Galactic early
B supergiants, including HD~37128, HD~38771 and HD~14956. For the 
B0--0.5\,Ia supergiants CNO
elemental abundances are consistent at the 0.2 dex level between the two
approaches, except that Urbaneja (2004) obtains a 0.25 dex lower N abundance 
for HD~37128 and a 0.3 dex lower C abundance for HD~38771. For HD~14956 
(B1.5\,Ia), the agreement is poorer - Urbaneja (2004) obtains a 0.55 dex 
lower C  abundance, 0.15 dex lower N abundance and 0.4 dex higher O 
abundance. 
Such differences should be taken into account regarding the {\it absolute} 
uncertainty in abundances resulting from such studies.

\begin{figure}[htbp]
\begin{center}
\includegraphics[width=0.9\columnwidth,clip,angle=0]{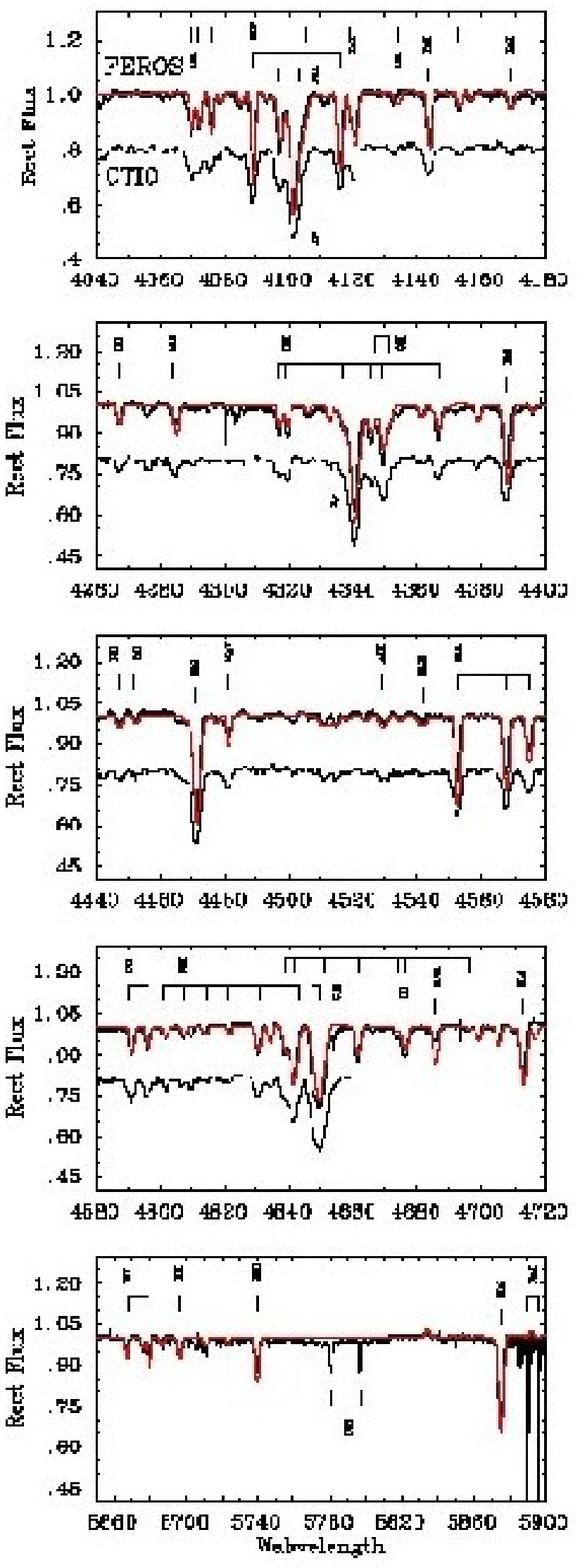}
\end{center}
\caption{Spectral comparison between FEROS spectroscopy of HD~115842 (solid black)
and synthetic spectroscopy (red in electronic version, dotted in paper version), 
with CTIO spectroscopy additionally
included, offset by -0.2 continuum units.}
\label{feros}
\end{figure}

 \section{Wind properties}

Recently,  Trundle et al. (2004) and Trundle \& Lennon (2005) have analysed a sample of 
SMC B supergiants using FASTWIND, based upon high and intermediate dispersion optical 
spectroscopy. Before we compare the wind properties of Galactic and SMC B supergiants
we wish to investigate whether our use of intermediate dispersion datasets introduces
systematic effects on derived stellar properties. 

\subsection{Analysis of high resolution observations of HD~115842}

HD~115842 was observed with both the CTIO 1.5m/Cassegrain and ESO 2.2m/FEROS, allowing
us to compare the derived properties of this star from the two datasets.
In fact, we were able to match our intermediate dispersion model for HD~115842 to the FEROS
echelle dataset, except that the  turbulence was increased from $v_{t}$= 10 km\,s$^{-1}$ to 
25 km\,s$^{-1}$. We compare  selected regions of the FEROS dataset and synthetic spectrum in 
Fig.~\ref{feros}, including  regions not covered by the CTIO dataset. Overall, 
the agreement is excellent, such that we are confident in the use of 
intermediate dispersion for our analysis.

 \subsection{Wind momenta for Galactic versus SMC B Supergiants}

\begin{figure}[htbp]
\begin{center}
\includegraphics[width=1\columnwidth,clip,angle=0]{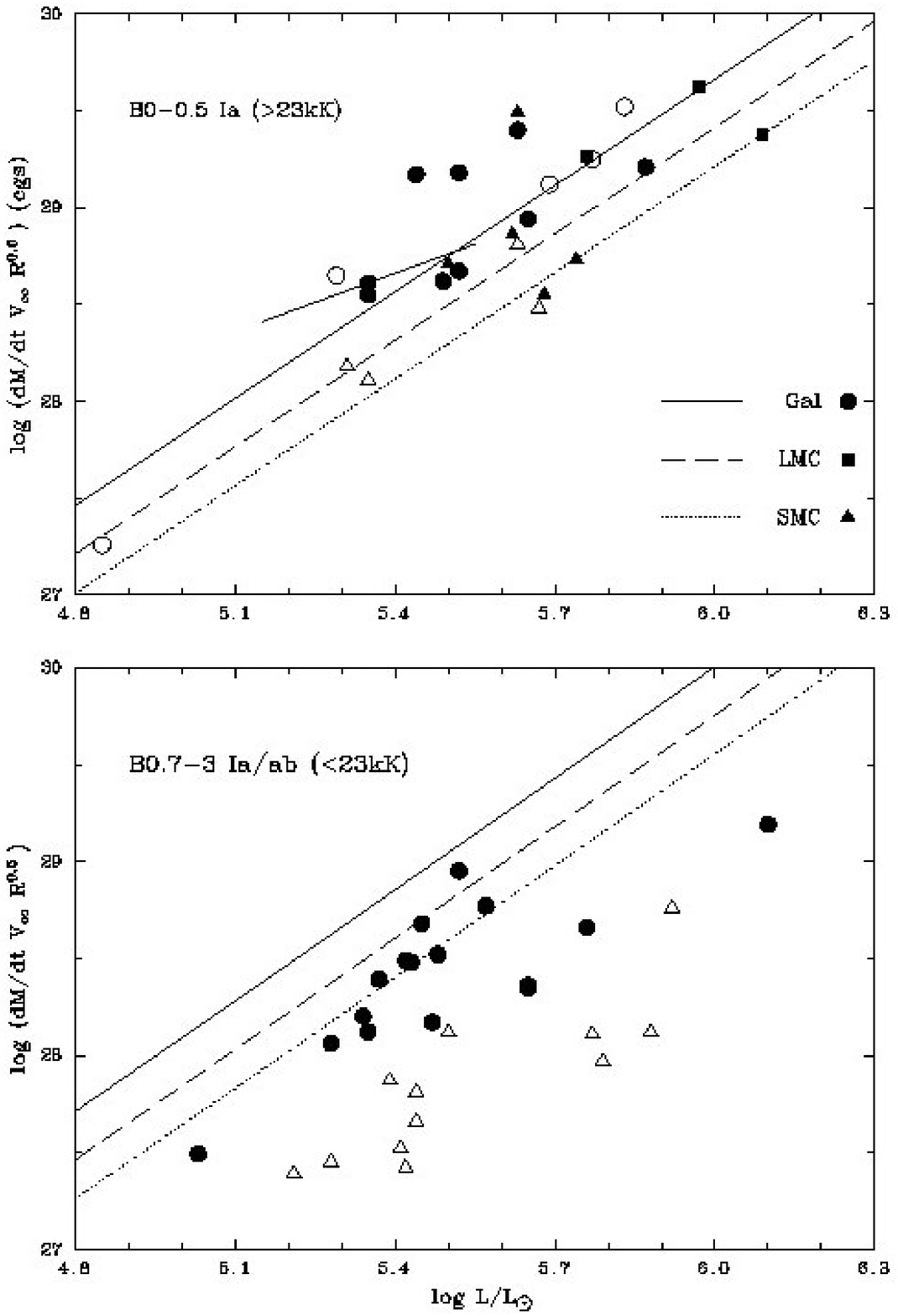}
\end{center}
\caption{Comparison between the Galactic B supergiant wind momenta presented here, 
separated between `early' ($>$23kK) and `mid' ($\leq$23kK) subtypes,  together
with other recent studies based on CMFGEN (filled, Evans et al. 2004b) or FASTWIND (open,
Trundle et al. 2004; Trundle \& Lennon 2005; Repolust et al. 2004; Herrero et al. 2002),
 plus radiatively driven wind theory
predictions by Vink et al. (2000, 2001) for early and mid B supergiants. The effect of an
uncertainty of $\pm$0.5 mag in distance modulus is indicated for HD~38771 (B0.5\,Ia).}
\label{wmr}
\end{figure}

We present the wind momenta of our Galactic sample versus luminosity  in Fig.~\ref{wmr},
separated between early B ($>$23kK) and mid B ($\leq$23kK) subtypes, 
together with current predictions for early and mid subtypes from Vink et al. (2000). 
In addition, we include  recent Magellanic Cloud 
results of Evans et al. (2004b), Trundle et al. (2004) and Trundle \& Lennon (2005).
The effect of an uncertainty in distance modulus of $\pm$0.5 mag is indicated for HD~38771.
This typical uncertainty 
illustrates why we do not attempt a calibration of the wind momentum luminosity relationship
at this time.

For B0--0.5 subtypes, Galactic wind momenta
lie close to the Vink et al. (2000) predictions for the majority of cases.
In three cases, the derived
wind momenta fall 0.5 dex below predictions, namely HD~91943, 115842 and 152234.
Due to this spread, there is no clear distinction between 
the observed wind momenta of early B supergiants 
in the SMC and Milky Way (compare triangles to 
circles), although this could be due to
small number statistics. If winds of early B supergiants
are clumped in the H$\alpha$ line forming region, we would need to shift the observed wind
momenta to smaller values.

For B0.7--3 subtypes, the measured mass-loss rates now lie typically
$\sim$0.5 dex below the calibration of Vink et al., 
with the exception of HD~2905 (BC0.7\,Ia). As above, if H$\alpha$ derived 
mass-loss rates of OB stars need to be 
corrected for clumped winds, as suggested by 
recent observational evidence (Massa 
et al. 2003; Evans et al. 2004b), the difference
between theory and observation  would be  exacerbated since the observed data 
points would move to lower values. In contrast with the case for the
earliest B supergiants, we
now see a clear separation between Galactic (circles) and SMC (triangles) 
B0.7--3 supergiants at approximately the predicted offset.

\subsection{Bistability jump}

Let us now return to the question of the bistability jump. 
First, having revised the
temperature scale for early B supergiants, let us investigate the 
sharp $\sim$21kK bistability 
jump of Lamers et  al. (1995) based on our present results. 

\begin{figure}[htbp]
\begin{center}
\includegraphics[width=1\columnwidth,clip,angle=0]{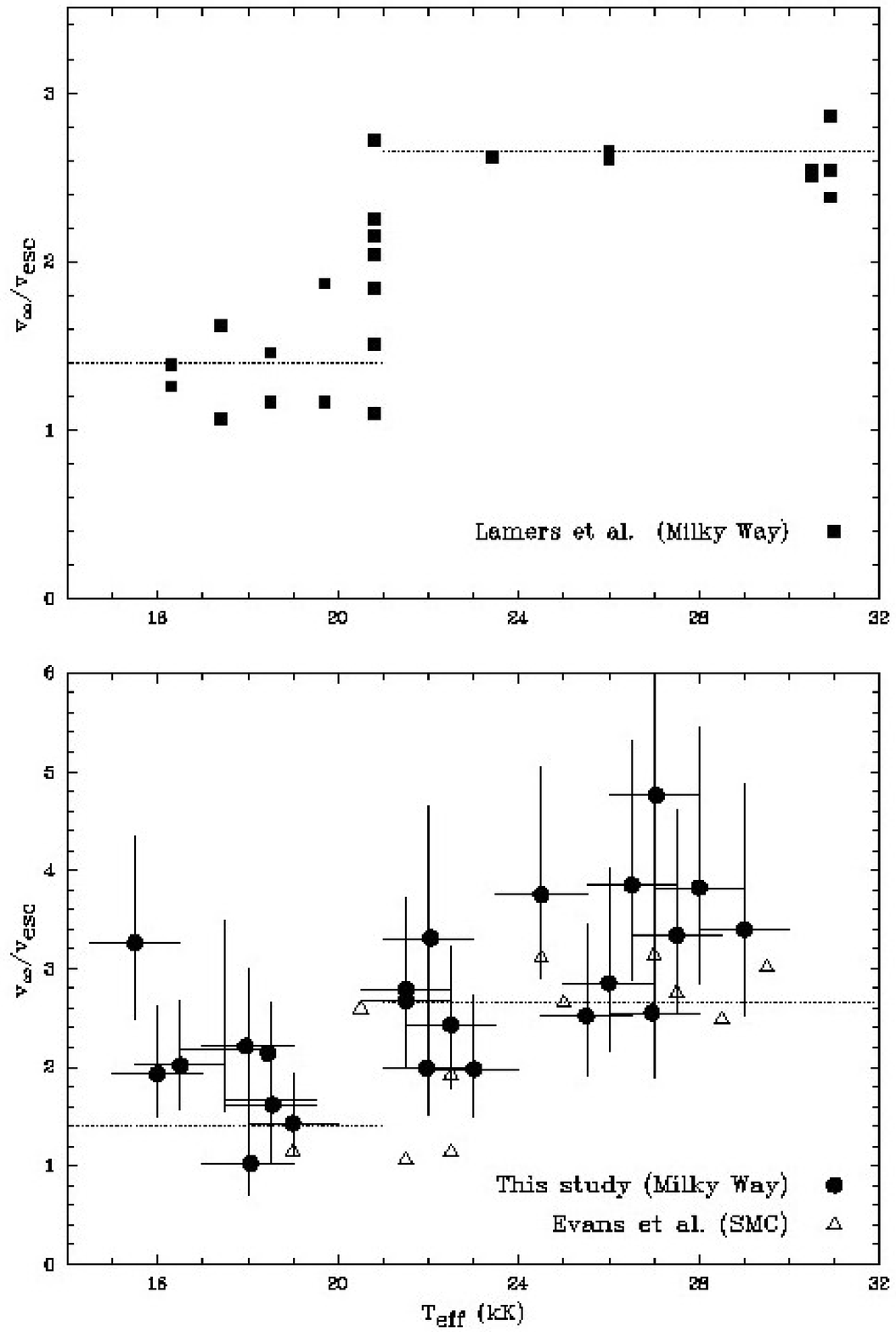}
\end{center}
\caption{The ratio $v_{\infty}/v_{\rm esc}$ as a function of 
effective temperature for {\bf (upper panel:)} Galactic B supergiants from Lamers et al.
(1995) together with the Kudritzki \& Puls (2000) scalings (dotted lines);
{\bf (lower panel):} Galactic B supergiants from the present study (filled circles),
together with results from Evans et al. (2004a, open triangles) 
for SMC B supergiants. Error bars relate to uncertainties in temperature ($\pm$1kK),
gravity ($\pm$0.15 to 0.2 dex) and absolute magnitude ($\pm$0.3 to 1 mag). We 
have omitted the two stars
 (HD~14956, HD~194279) 
for which reliable wind velocities are not known.
\label{bistab}}
\end{figure}

In the upper panel of Fig.~\ref{bistab}
we repeat the original Lamers et al. (1995) O9.5--B3 supergiant sample, 
which illustrates the sharp 21kK jump. This clear step is due to Lamers
et al. adopting a single temperature of 20.8kK for all their seven B1 
supergiants. In the lower panel of Fig.~\ref{bistab} we present our current  results,
where we have omitted HD~14956 and HD~194279  for which no reliable wind
velocities are known.  This sample includes 
several sources in the original Lamers et al. study,
together with  values from Evans et al. (2004a), based
on measured HST/STIS wind velocities plus  model atmosphere results from Evans
et al. (2004b), Trundle et al. (2004), Trundle \& Lennon (2005). 

Error bars allow for uncertainties in distance, surface gravity and temperature.
Since distances to Galactic early-type supergiants are notoriously
imprecise (witness the Hipparcos versus Ori OB1 distance estimates to HD~38771),
we provide error bars allowing for uncertainties in distance and surface gravity.
Of these, the latter dominates, even for cases with particularly poorly constrained
distances, such as HD~190603.

In contrast with the earlier study, there is a gradual downward trend of 
$v_{\infty}/v_{\rm esc}$ with temperature, albeit with a large scatter. 
For $T_{\rm eff}>24$kK (approximately B0.5\,Ia and earlier), 
$v_{\infty}/v_{\rm esc} \sim 3.4$, 
for 20kK$\leq T_{\rm eff} \leq$24kK (approximately B0.7--1\,Ia) 
$v_{\infty}/v_{\rm esc} \sim 2.5$,  and for $T_{\rm eff}<$20kK 
(approximately B1.5\,Ia and later) $v_{\infty}/v_{\rm esc} \sim 1.9$. 
This reveals that the B1 `jump' is misleading in the context of 
normal B  supergiants. 
Prinja \& Massa (1998) came to similar conclusions based on a larger B 
supergiant sample, albeit with an adopted subtype-temperature calibration.
Amongst the present sample,
the hypergiant HD~152236 (B1.5\,Ia$^+$) has the lowest ratio
(1.0$^{+0.6}_{-0.3}$) whilst HD~37128 (B0\,Ia) has the highest ratio of all (4.8$^{+2.0}_{-1.2}$).

Vink et al. (1999) predicted a dramatic increase in wind densities of early B supergiants
below $\sim$25kK due to Fe, the dominant line driving ion, shifting to lower ionization stage.
This temperature is close to the boundary between B0.5 and B0.7 subtypes where changes in
$v_{\infty}/v_{\rm esc}$ 
do occur, although the original Lamers et al. (1995) 21kK boundary is an equally 
valid $v_{\infty}/v_{\rm esc}$ step. However, no significant increase in mass-loss rate below $\sim$24kK is 
identified from our results.

Finally, recall empirical values of the Lamers et al. (1995) `bistability jump' 
were adopted by Vink et al. (1999, 2000) in their radiatively driven wind calculations.
These predictions have subsequently been used in evolutionary calculations (e.g. Meynet \& Maeder
2000) and spectral synthesis calculations (e.g. Rix et al. 2004). Consequently, 
our re-determination of  physical parameters and wind properties of early B supergiants has 
potential consequences for evolutionary and spectral synthesis calculations.

\subsection{UV morphology}

The primary role of the present study is to provide 
physical parameters and wind properties of Galactic B supergiants 
based on optical 
diagnostics, with reference to the bistability `jump'. We shall
postpone a detailed discussion of UV spectral comparisons to elsewhere, 
but wish to comment on one specific aspect relating to UV morphology.
Walborn (1971) introduced the B0.7 subtype on the basis
of the behaviour of Si\,{\sc iii} $\lambda$4552/Si\,{\sc iv} $\lambda$4089.
 Walborn \& Nichols-Bohlin  (1987) found that
B0.5\,Ia and B0.7\,Ia supergiants were well defined and quite distinct in
their UV spectral morphology, with regard to the presence/absence 
of C\,{\sc ii} and Al\,{\sc iii} P Cygni lines, plus the shape of the 
Si\,{\sc iv} absorptions.  
Between B0.5 and B0.7  the blue-shifted 
Si\,{\sc iv} absorptions suddenly  become narrower and deeper, the latter 
reversing the smooth trend from earlier types.

\begin{figure*}[htbp]
\begin{center}
\includegraphics[width=1.8\columnwidth,clip,angle=0]{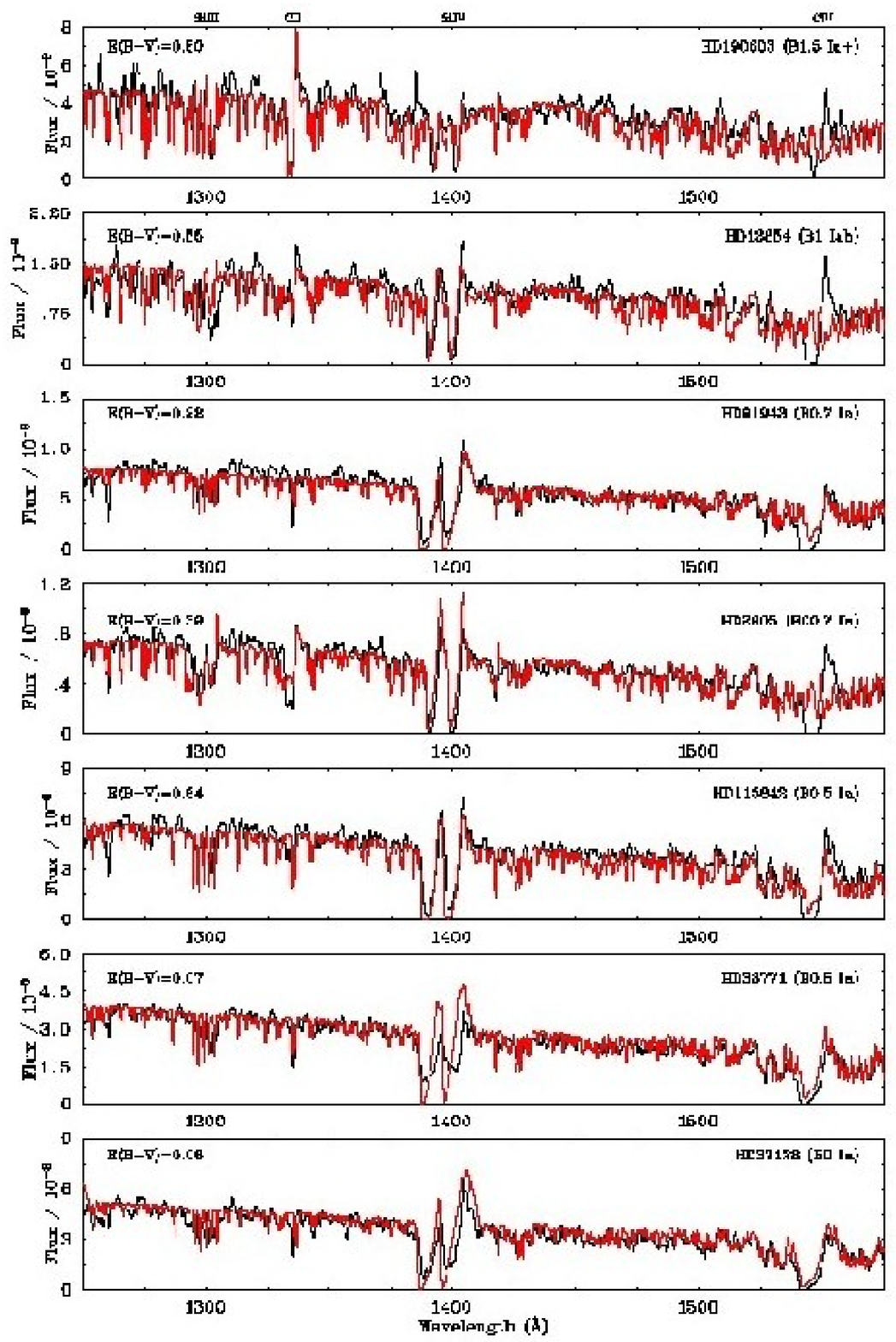}
\end{center}
\caption{{\bf (a)} Comparison between de-reddened HIRES IUE spectroscopy ($\lambda\lambda$1250--1575)
of selected early B supergiants (solid black) and synthetic UV spectra resulting from 
optical analyses (red in electronic version, dotted in paper version).\label{fuv}}
\end{figure*}

\addtocounter{figure}{-1}

\begin{figure*}[htbp]
\begin{center}
\includegraphics[width=1.8\columnwidth,clip,angle=0]{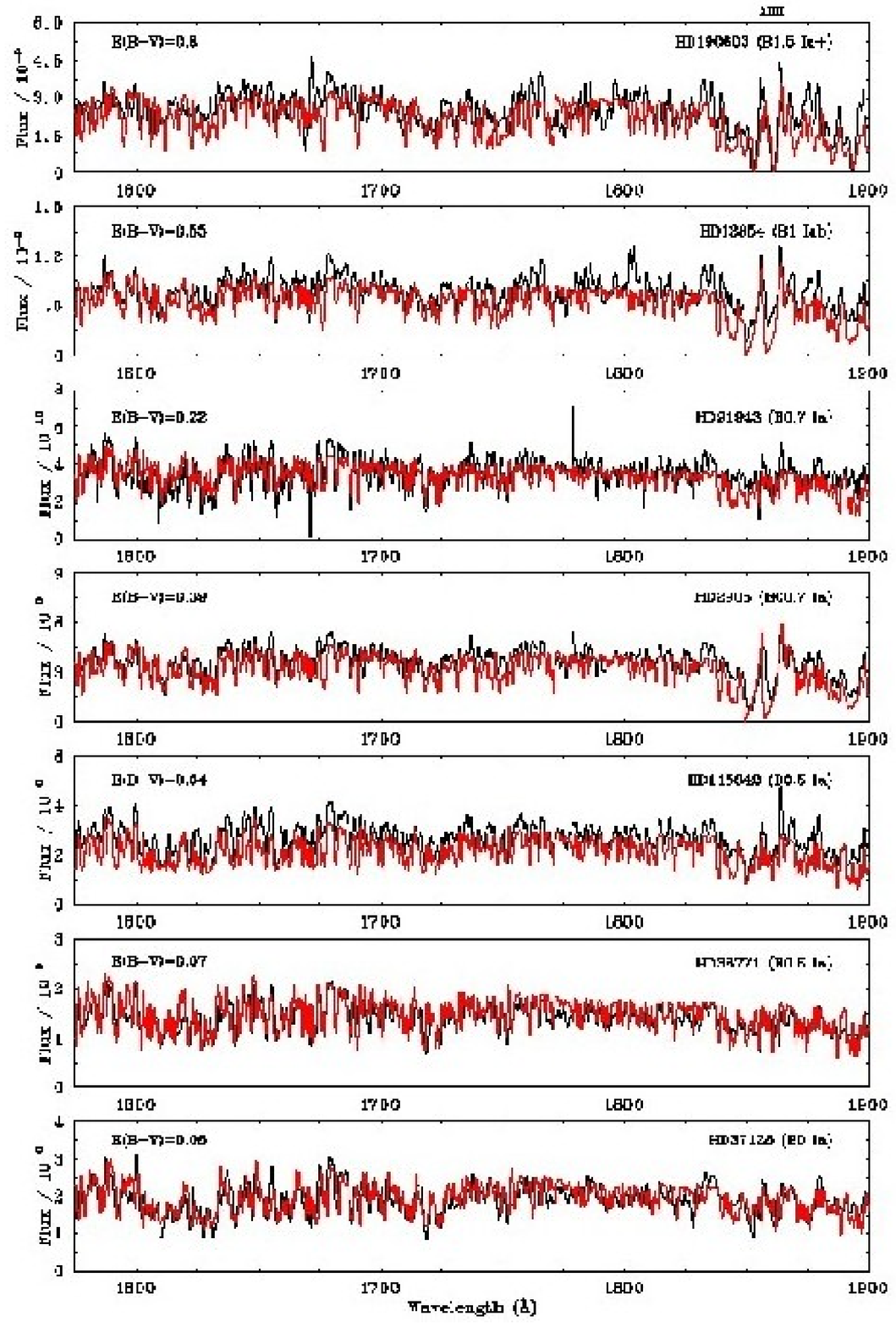}
\end{center}
\caption{{\bf b)} Comparison between de-reddened HIRES IUE spectroscopy ($\lambda\lambda$1575--1900)
of selected early B supergiants (solid black) and synthetic UV spectra resulting from 
optical analyses (red in electronic version, dotted in paper version).\label{nuv}}
\end{figure*}

We have shown that the bistability `jump' does not have
a well defined edge, but rather a general trend to lower $v_{\infty}/v_{\rm 
esc}$, particularly between B0.5 and B0.7 subtypes. How successful are 
our models, which are based on optical diagnostics, at reproducing these 
distinct UV properties?

In Figs.~\ref{fuv}(ab) we present comparisons 
between de-reddened IUE (HIRES short wavelength SWP) 
observations of selected B supergiants and  synthetic spectra. The UV 
spectra of B supergiants are greatly affected by metal blanketing.
Indeed, $\sim$95\% of the strong ($W_{\lambda} \geq$ 0.01\AA) lines in the 
$\lambda\lambda$1250--1900 region of early B supergiants 
are due to Fe\,{\sc iii-iv} transitions. Fe\,{\sc iv} dominates the far-UV
blanketing at B0, equal numbers of Fe\,{\sc iii} and Fe\,{\sc iv} lines contribute
at B0.7, whilst Fe\,{\sc iii} dominates at later subtypes. The density of spectral
lines is so large that there are no clean UV continuum regions, such that
our comparisons are between fluxed datasets. 

In {\it general}, the UV spectra are fairly  well reproduced
by the optically derived diagnostics, both prominent P Cygni wind 
profiles of Si\,{\sc iv}, C\,{\sc iv} and Al\,{\sc iii}, 
and the primarily Fe-blanketed continuum. The most notable exceptions are
(i) C\,{\sc iv}  $\lambda\lambda$1548--51 for subtypes later than B0.5
for which negligible P Cygni emission is predicted, in 
dramatic contrast with observation -- N\,{\sc v} $\lambda\lambda$1238--42
suffers a similar problem for all subtypes -- and (ii)  The P 
Cygni absorption  
strength  of Si\,{\sc iv}   $\lambda\lambda$1393--1402 is predicted too 
strong in B0--0.5 supergiants.

Observationally, C\,{\sc iv}  and C\,{\sc ii} lines co-exist in B0.7--B2 
supergiants, whilst the the former is effectively absent for stellar 
temperatures below  $\sim$25kK. One obvious means of maintaining high 
ionization states at relatively low stellar temperatures (and hence
C\,{\sc iv}, N\,{\sc v} wind lines) is via X-rays,  
produced  by shocks in the stellar wind, 
which are also responsible for the clumped winds discussed above. 
Quantitatively addressing this question is beyond the scope of the
present paper, and will be investigated in a future study.

Nevertheless, we also have a partial answer to  the reversal of the 
conventional UV morphologies for two early B  supergiants, namely 
the B0.5\,Ia supergiant HD\,115842 and B0.7\,Ia supergiant HD\,91943 
(Walborn et al. 1995). In general, B0.5  and B0.7 supergiants are 
distinct in terms of  C\,{\sc ii} and Al\,{\sc iii} due to the 
substantial differences in ionization between $\sim$26kK (B0.5) and 
$\sim$22.5kK (B0.7) of these elements.  Indeed, the unusually high stellar 
temperature of $\sim$24.5kK for HD~91943 (B0.7\,Ia) implies a
high degree of ionization, such that negligible wind features are 
predicted, or observed, at  C\,{\sc ii} and/or Al\,{\sc iii}. This high 
temperature results from an  unusually low wind density, relative to other 
B0.7 supergiants, as  evidenced by its  weak H$\alpha$ signature in 
Fig.~\ref{fig3}).   In contrast, HD~115842 (B0.5\,Ia) has a rather strong
wind, and so a low stellar temperature, with the Al\,{\sc iii} feature 
seen weakly.

Finally, we turn to the other morphological difference between B0.5
and B0.7 supergiants in the UV, i.e. narrower and deeper 
Si\,{\sc iv} absorption at later subtypes. 
 We have demonstrated above
that  there is indeed a trend for  stars with   $T_{\rm eff} \geq 24$kK 
-- approximately dividing B0.5 from B0.7 subtypes 
-- to possess higher  $v_{\infty}/v_{\rm esc}$  ratios, i.e. broad,
shallow Si\,{\sc iv} absorption is observed in B0.5 stars with high
wind velocities, with narrow, deep Si\,{\sc iv} absorption observed in 
B0.7 stars with low wind velocities.

Omitting HD~91943 and HD~115842 from the sample of Howarth et al. (1997), the
average B0.5 and B0.7 Ia supergiant wind velocities are 1535 km/s and 955
km/s, respectively. The wind velocities of the morphologically 
unusual stars are indeed unusual. HD~91943 (B0.7\,Ia) has a
fast wind of 1470 km/s, whilst HD~115842 (B0.5\,Ia) has a
slow wind of 1180 km/s. Of course, a physical explanation is
incomplete since the present models fail to predict the observed
behaviour of Si\,{\sc iv} in cases with fast winds amongst B0--B0.5
supergiants, i.e. all except HD~115842 show too strong Si\,{\sc iv} 
absorption.  

Clumping may play a role here too. If the conventional mass-loss
diagnostics (e.g. H$\alpha$) in B  supergiants formed in regions where
the wind is clumped, the global mass-loss rate will decrease, potentially
desaturating Si\,{\sc iv} absorption. Similar conclusions were reached
by Evans et al. (2004b) in their far-UV study of late O and early B 
supergiants in the Magellanic Clouds, in which the observed
S\,{\sc iv}  $\lambda\lambda$1062--73 absorption could only be matched
by clumped models (see also Prinja et al. 2004).  A test calculation
has been carried out for HD~38771 in which clumping was included, with a 
volume filling factor of 0.1. Si\,{\sc iv} $\lambda$1393--1402 was not found 
to  de-saturate in this case (C\,{\sc iv} $\lambda$1548-51 did
de-saturate), 
indicating either a higher clumping factor or an alternative explanation.

\section{Conclusions}\label{sect5}

We present detailed optical studies of a sample of Galactic B supergiants
in order to determine physical parameters and wind properties based on contemporary
model atmosphere techniques, for comparison with recent results on Magellanic
Cloud B supergiants (e.g. Evans et al. 2004b, Trundle et al. 2004). With
respect to Kudritzki et al. (1999) who carried out a similar study based on
an adopted temperature scale and a  H$\alpha$ mass-loss
determination from an unblanketed code, we obtain lower stellar temperatures (and hence
luminosities) and either comparable (early B) or higher (mid B) mass-loss rates.
CNO elements are found to be partially CNO processed in general. On average,
[N/C] and [N/O] are increased relative to (recently revised) Solar abundances
by +1.1 dex and +0.8 dex, respectively. The morphologically unusual B supergiant
HD~2905 (BC0.7\,Ia) indeed has an unusually low N enrichment and C/O depletion.

With
respect to recent theoretical predictions, we find reasonable agreement
with those of Vink et al. (1999, 2000) for early ($\geq$23kK) B supergiants
assuming homogeneous winds, although the predicted winds for later subtype 
B are too strong. This would only be exacerbated in the case of 
clumped winds in the H$\alpha$ line forming region (Evans et al. 2004b).
With respect to recent SMC B supergiant studies,  Galactic mid-B supergiants
 are observed to have systematically stronger winds, 
although the situation is less clear for early B supergiants.
We obtain reasonable agreement between the mid-IR continuum derived mass-loss rates 
of Barlow \& Cohen (1977) and H$\alpha$ mass-loss rates suggesting similar clumping
factors for these regions, in contrast with expectation (Runacres \& Owocki 2002).

We have investigated the so-called `bistability jump' amongst early B supergiants,
in which Lamers et al. (1995) claimed a step in the ratio 
$v_{\infty}/v_{\rm esc}$ at 21kK (B1) from 2.65 to 1.3. From our detailed study
of  larger sample close to this subtype, we find that there is in fact a gradual
decline in $v_{\infty}/v_{\rm esc}$ from early to mid B supergiants (also identified
by Evans et al. 2004a for SMC supergiants) from $\sim$3.4 above 24kK, 
$\sim$2.5 for 20--24kK, and $\sim$1.9 below 20kK.

We compare our optically derived spectral fits to UV spectrophotometry of early B
supergiants. Morphological differences in the UV between B0.5 ($\sim$26kK)
and B0.7 ($\sim$23kK) subtypes (Walborn et al. 1995) are attributed to the difference in 
ionization  and wind velocity between these subtypes. 
Recall Vink et al. (1999) predicted a
 shift in  ionization around 25kK for B supergiants, due to which wind densities 
increase dramatically below  this temperature. The former does indeed match the observed 
shift in wind  velocities at $\sim$24kK, although no 
significant increase in  mass-loss rate is inferred from our results,  affecting evolutionary
models which make use of such predictions (e.g. Meynet \& Maeder 2003).
Finally, we discuss potential solutions to spectroscopic discrepancies at Si\,{\sc iv} (at 
B0--0.5) and C\,{\sc iv} (at B0.7--1.5) which again suggest the presence of shocked, clumped
winds in the line forming regions. Further progress requires the development of spectroscopic
clumping diagnostics.


\begin{acknowledgements}
We wish to thank John Hillier for providing CMFGEN to the general astronomical community.
PAC and DJL thank the STScI for financial support from the
Director's Discretionary Research Fund  where this work was initiated. 
Southern targets were observed at the CTIO 1.5m telescope by Sergio Gonzalez Huerta
via the SMARTS Consortium. STScI participation in SMARTS is
also funded by the DDRF. We are grateful to Steve Smartt for providing high S/N WHT optical spectroscopy
for several targets, and Raman Prinja for providing digital max/min H$\alpha$ spectroscopy of $\epsilon$ Ori. 
Chris Evans kindly obtained the FEROS spectrum of
HD~115842 on our behalf, to whom we are grateful. Finally, we appreciate useful comments on this
paper from Jo Puls and from an anonymous referee.
\end{acknowledgements}

\end{document}